\documentclass[12pt]{report}

\usepackage{ODUthesisFall2016}
\usepackage{indentfirst}
\usepackage{cite}
\usepackage[table]{xcolor}
\usepackage{array,booktabs,ragged2e}
\newcolumntype{P}[1]{>{\centering\arraybackslash}p{#1}}
\newcolumntype{M}[1]{>{\centering\arraybackslash}m{#1}}
\newcolumntype{R}[1]{>{\raggedleft\arraybackslash}m{#1}}
\usepackage[pdfauthor={John A. Snoap},
			pdftitle={Physical Layer Secret Key Agreement Using One-Bit Quantization and Low-Density Parity-Check Codes},
			pdfsubject={Physical Layer Secret Key Agreement Using One-Bit Quantization and Low-Density Parity-Check Codes},
			pdfkeywords={Physical Layer Secret Key Agreement}]{hyperref}
\usepackage[labelsep=period, labelfont=bf]{caption} 
\hypersetup
{
    colorlinks=true, 
    linktoc=all,     
    linkcolor=black, 
    allcolors=black, 
    bookmarksopen=true,
    bookmarksnumbered=true,
    pdfstartview=FitH,
    pdfremotestartview=FitH,
}
\usepackage{bookmark}

\usepackage [english]{babel}
\usepackage [autostyle, english = american]{csquotes}
\MakeOuterQuote{"}

\usepackage[titles]{tocloft}
\cftsetindents{figure}{0em}{2.55em} 
\cftsetindents{table}{0em}{2.55em} 

\usepackage{graphicx}
\graphicspath{{./figures/}}
\DeclareGraphicsExtensions{.pdf,.eps}
\usepackage{epstopdf}

\usepackage{amsmath}
\usepackage{algpseudocode}
\newcounter{corol}
\renewcommand{\thecorol}{\arabic{section}.\arabic{corol}}

\newcounter{lemma}
\renewcommand{\thelemma}{\arabic{lemma}}

\newcounter{theorem}

\newcommand{\kv}{\mathbf{k}}

\newcommand{\pv}{\mathbf{p}}

\newcommand{\be}{\begin{equation}}
\newcommand{\ee}{\end{equation}}

\begin{document}

\title{Physical Layer Secret Key Agreement Using One-Bit Quantization and Low-Density Parity-Check Codes}

\author{John A. Snoap}
\principaladviser{Dimitrie C. Popescu}
\member{Dean J. Krusienski}
\member{W. Steven Gray}

\degrees{B.S. April 2014, Oklahoma Christian University}

\dept{Electrical \& Computer Engineering}          

\submitdate{December 2016}

\phdfalse



\vita{John Andrew Snoap was born January 7, 1992, in Orlando, Florida.  Because of his father's position in the military, John lived in many different states, but finished growing up in Chesapeake, Virginia.  After graduating from Hickory High School in Chesapeake, Virginia in 2010, he went on to earn his Bachelor of Science degree in Electrical Engineering from Oklahoma Christian University in April 2014.  While earning his bachelor's degree, John worked two summer internships at Newport News Shipbuilding (NNS) in Newport News, Virginia.  After earning his bachelor's degree, John began working for NNS, where he is currently employed full-time.  In the Fall of 2014, he enrolled at Old Dominion University as a part-time student, earning his Master of Science degree in Electrical and Computer Engineering in December 2016.}

\abstract{\noindent Physical layer approaches for generating secret encryption keys for wireless systems using channel information have attracted increased interest from researchers in recent years.  This paper presents a new approach for calculating log-likelihood ratios (LLRs) for secret key generation that is based on one-bit quantization of channel measurements and the difference between channel estimates at legitimate reciprocal nodes.  The studied secret key agreement approach, which implements advantage distillation along with information reconciliation using Slepian-Wolf low-density parity-check (LDPC) codes, is discussed and illustrated with numerical results obtained from simulations.  These results show the probability of bit disagreement for keys generated using the proposed LLR calculations compared with alternative LLR calculation methods for key generation based on channel state information.  The proposed LLR calculations are shown to be an improvement to the studied approach of physical layer secret key agreement.}

\beforepreface

\prefacesection{Acknowledgements}
I would like to acknowledge and express my gratitude to Dr. Dimitrie C. Popescu for being both my constructive advisor and, foremost, my course instructor.  His foresight has enabled me to complete pertinent courses in an appropriate order, even though I was a part-time student and was fully employed.  Also, in all four courses in which he was my instructor, I always learned new and interesting information concerning communications.  Through his guidance, I was able to perform research while working as a team with his Ph.D. student, Sayg{\i}n Bak\c{s}i.  The knowledge, support, patience, and flexibility of both Dr. Popescu and his student, Sayg{\i}n, were key to the completion of my graduate work.

I would like to thank Dr. Dean J. Krusienski and Dr. W. Steven Gray for serving on my thesis committee.  Through the courses in which they instructed me, I was pointed in an encouraging direction, which helped lead to the completion of my graduate work.  I asked them to serve as committee members because of how much I learned from them and the courses they instructed.  I am grateful that they accepted.

I would like to thank Newport News Shipbuilding (NNS) for reimbursing me for the cost of tuition and books.  I would also like to thank my supervisor and colleagues at NNS for their flexibility in allowing me to attend on-campus classes and complete my coursework.

Finally, I would like to thank my parents for encouraging me and instilling in me the joy of putting Christ first, others second, and myself last.  It is this motivation that keeps me striving to do the best that I can with the talents and opportunities that have been given to me.
\afterpreface

\chapter{Introduction}\label{sec:intro}
Wireless communication systems, along with the services they provide, have become an essential component of modern society.  Due to the shared nature of the transmission medium, the radio frequency (RF) spectrum, wireless communications are inherently insecure and are prone to eavesdropping.  To protect against eavesdropping and to ensure the confidentiality of transmitted data, many wireless systems employ encryption, using secret keys available only to the transmitter and the corresponding legitimate receiver.

Agreeing upon what secret key to use for encryption is relatively simple when a separate secure channel is in place.  For example, if one were to configure a home network router's Wi-Fi, one could manually set the Wi-Fi password on the router, and then manually enter the Wi-Fi password onto other Wi-Fi enabled devices that connect to it.  In this scenario, the person setting the network password acts as the separate secure "channel" being used to share the secret key between the network router and the Wi-Fi enabled devices.

However, when the only channel available to share a secret key is insecure, guaranteeing agreement between the secret keys at transmitter and receiver is challenging.  The agreement must be accomplished such that legitimate users attain the secret key, while eavesdroppers are unable to ascertain the secret key.  For example, if Alice and Bob want to communicate secretly over a wireless channel, but they cannot come close enough to each other to exchange a secret key without an eavesdropper overhearing, then they will need to use a different method to securely transfer a secret key.

A common method of securely exchanging secret keys over insecure channels is to employ public-key cryptosystems, such as Rivest-Shamir-Adleman (RSA) \cite{RSA78}.  
While public-key cryptosystems are currently computationally secure, they are not unconditionally secure.  In recent years, to strive towards unconditional secrecy, various physical layer approaches have been proposed for generating encryption keys based on channel state information \cite{lai_etal_keygen2013}.  Improvements to these theoretical approaches must be accomplished in order to bring unconditionally secure communications closer to a practical reality.  Therefore, exploring and improving these theoretical approaches is important and has been of increased interest for researchers in recent years.

\newpage

\section{Wireless Security}\label{sec:wirelessSecurity}
Privacy and security in communications were not always of utmost importance at the beginning of wireless communications.  Simply successfully transmitting information from a transmitter to a receiver was a more important goal.  Some of the first methods of wireless transmission consisted of amplitude and frequency modulated carrier signals which used narrowband transmission techniques.  However, these narrowband transmission techniques were not secure, because any receiver within range of the transmitter could tune into, demodulate, and extract all the information from the wireless signal.

By World War II, radio communications had grown into a useful technology, and efforts were beginning to concentrate on issues of wireless security \cite{SSComms}.  It was in 1942 that a patent for a "Secret Communication System" was issued to Hedy Kiesler Markey, who was actually the young actress Hedy Lamarr, and George Antheil, who had previously synchronized player pianos using electronics \cite{HedyLamarr}.

\begin{figure}
\centering
\includegraphics[scale=0.75]{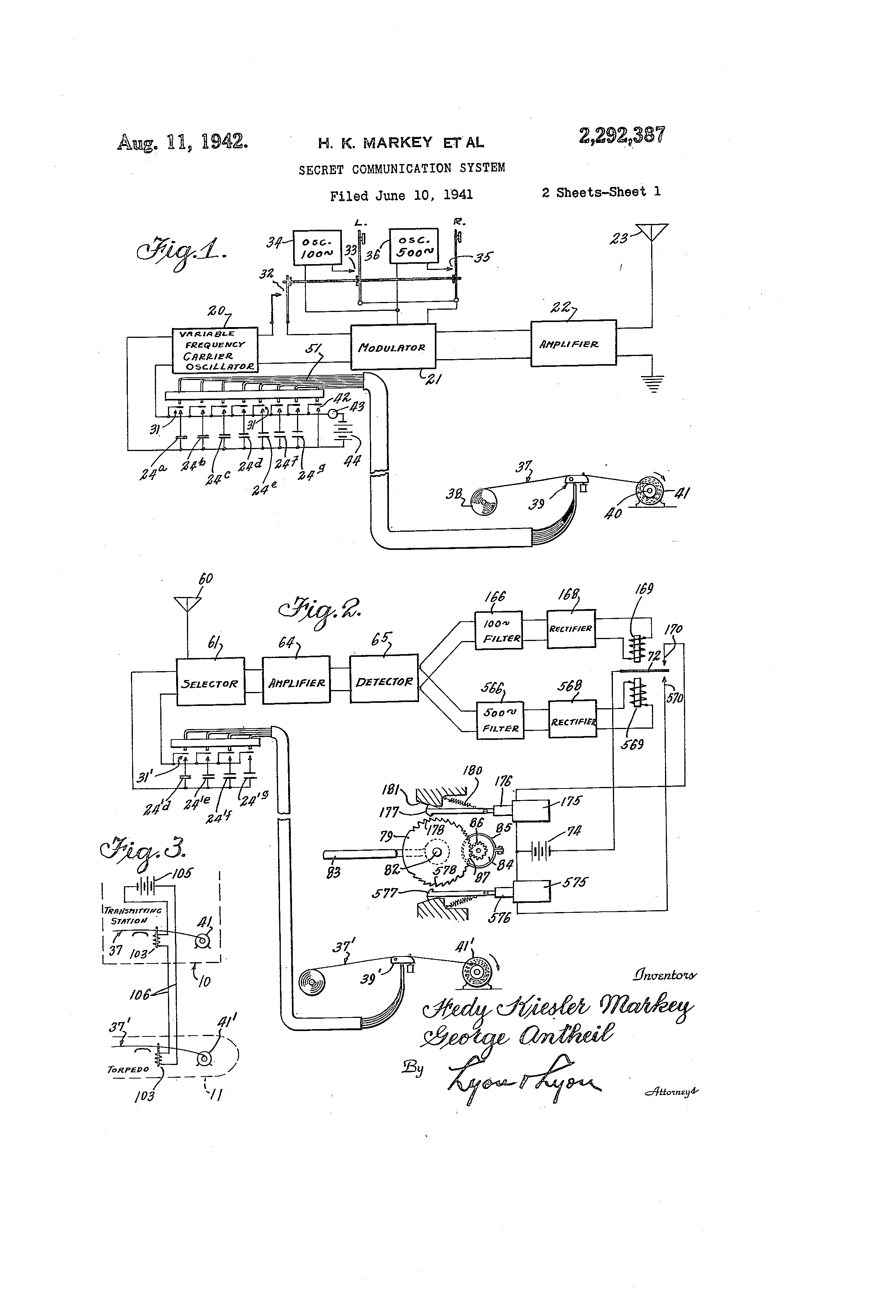}
\caption[The Player Piano Secret Communication System Sheet 1.]{\textbf{The Player Piano Secret Communication System Sheet 1.}}\label{fig:Hedy0}
\end{figure}

\begin{figure}
\centering
\includegraphics[scale=0.75]{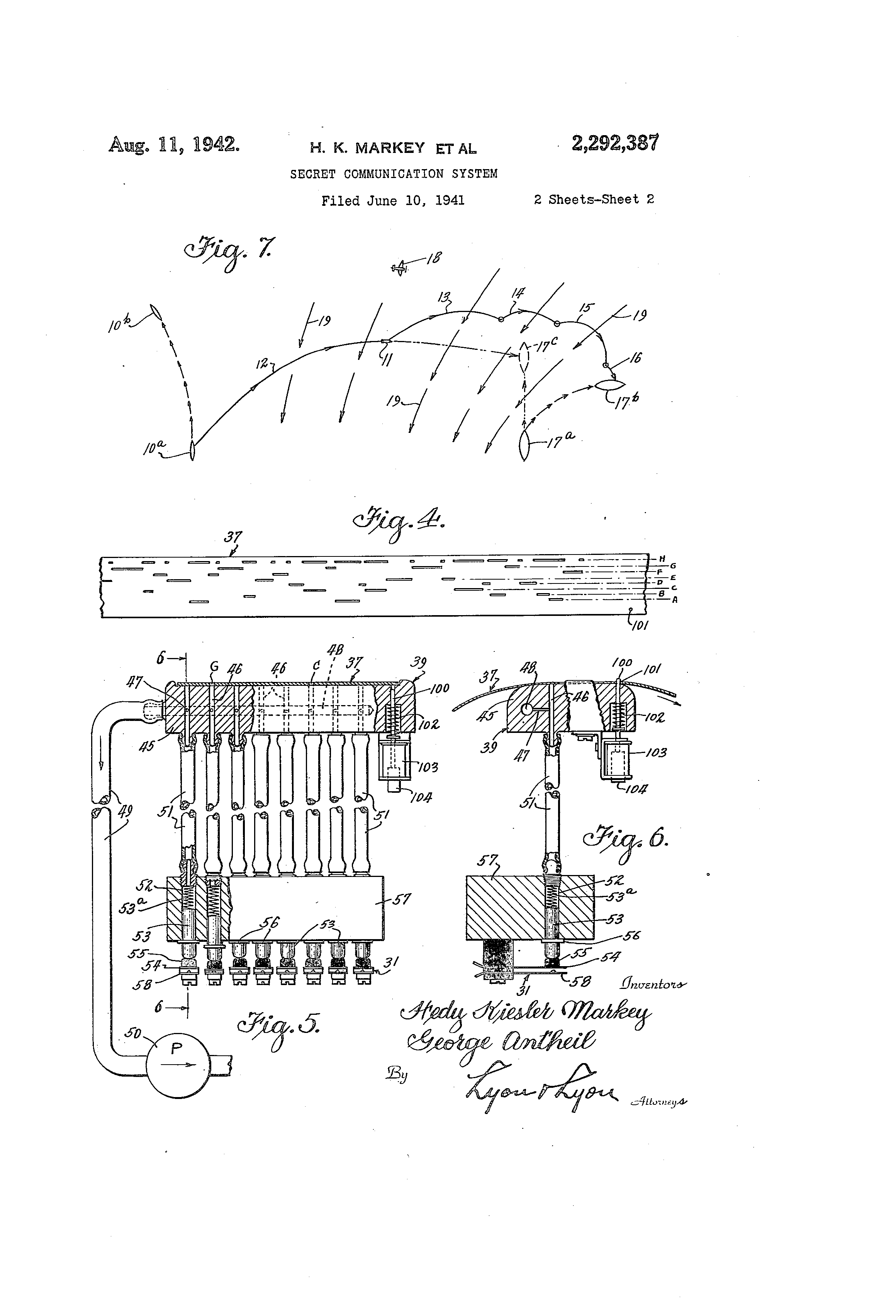}
\caption[The Player Piano Secret Communication System Sheet 2.]{\textbf{The Player Piano Secret Communication System Sheet 2.}}\label{fig:Hedy1}
\end{figure}

Together they invented an idea for a frequency-hopping system of radio control that used 88 keys (like a piano) for the frequencies it would use \cite{george1942secret}.  A transmitter would employ slotted paper rolls like those in player pianos, displayed in Figures~\ref{fig:Hedy0} and \ref{fig:Hedy1}, to determine a pseudo-random sequence and duration of signals in the 88 frequencies it would use.  The receiver would have an identical roll to receive the transmissions.  Without knowing the sequence, an enemy could neither detect the transmissions nor jam the transmissions.

While Lamarr and Antheil's frequency hopping scheme was made for the purpose of creating a "Secret Communication System", the Spread Spectrum techniques they pioneered had the ability to allow multiple users to access a communication channel.  It was their multiple access abilities that were used by Qualcomm.

Qualcomm developed their CDMA2000 mobile technology for third-generation (3G) cell phone communications in North America.  Code Division Multiple Access (CDMA) is a direct sequence (DS) type of spread spectrum multiple access (SSMA) technique.  In a spread spectrum technique, the bandwidth of the data is spread across many frequencies, so that each frequency is used evenly and with much less transmit power.  A pseudo-random noise code sequence converts a narrowband signal to a wideband noise-like signal.  The pseudo-random noise code sequence used in CDMA is what allows the communication technique to offer multiple users access to the same radio spectrum at once.  Each user has his or her own code word to generate a pseudo-random noise sequence for his or her communication.  Also, each code word used within the CDMA system is approximately orthogonal to all other code words.  That way, each user can only properly transmit and receive data associated with their code word to and from the cell tower.  All other transmissions formed using other code words are simply seen as noise.

While CDMA clearly allows multiple users to access the same frequency spectrum at the same time, the code word that is employed also makes the transmission scheme more secure than a typical narrowband radio transmission, because it is more difficult to detect transmissions that appear as noise.  In addition, only a person who has the code word can understand what is sent.  The code word ends up behaving in a similar manner to a secret key in cryptography.

\section{Cryptography}\label{sec:crypto}
Cryptography is the study of ways to disguise messages so as to avert unauthorized interception \cite[Ch. 14]{sklar2001digital}.  The terms encipher and encrypt refer to the message transformation performed at the transmitter to change plaintext (the original message) into ciphertext (the encrypted message).  The terms decipher and decrypt refer to the inverse transformation performed at the receiver to change the ciphertext back into plaintext.

Two fundamental reasons for using cryptosystems in communications are (1) privacy, to prevent unauthorized persons from eavesdropping and extracting information from the channel; and (2) authentication, to prevent unauthorized persons from spoofing and injecting information into the channel \cite{CTSS}.  To give an example of an instance in which authentication may be necessary, consider the scenario of a funds transfer or a contract negotiation.  If the transfer or negotiation occur electronically, it may be important to provide the electronic equivalent of a written signature in order to avoid or to settle any dispute between the sender and receiver as to what message, if any, was sent.

Generally, any cryptosystem has two major requirements \cite[Ch. 14]{sklar2001digital}:

\begin{enumerate}
\item To provide an easy and inexpensive means of encryption and decryption to all authorized users in possession of the appropriate key
\item To ensure that the cryptanalyst's task of producing an estimate of the plaintext without benefit of the key is made difficult and expensive
\end{enumerate}

Successful cryptosystems are classified as being either unconditionally secure \cite{CTSS} or computationally secure \cite{RSA78}.  A system is said to be unconditionally secure when the amount of information available to the cryptanalyst is insufficient to determine the encryption and decryption transformations, no matter how much computing power the cryptanalyst has available \cite[Ch. 14]{sklar2001digital}.  One unconditionally secure system, called a one-time pad \cite{OTP}, involves encrypting a message with a random key that is used one time only.  The key is never reused; hence the cryptanalyst is denied information that might be useful against subsequent transmissions because those subsequent transmissions use a different key.  Although the one-time pad system is unconditionally secure \cite{CTSS}, it has limited use in current communication systems, since a new key would have to be distributed for each new message --- a great logistical burden.  The logistical burden of distributing keys is one that could be lightened by using a physical layer approach as will be explained in section \ref{sec:physLayerKeyGen}.

Although some systems can be proven to be unconditionally secure, currently there is no known way to demonstrate security for an arbitrary cryptosystem \cite{RSA78}.  Hence, the specifications for most cryptosystems rely on the less formal designation of computational security for $x$ number of years, which means that under circumstances favorable to the cryptanalyst (i.e., using state-of-the-art computers) the system security could be broken in a period of $x$ years, but could not be broken in less than $x$ years \cite{RSA78}.

\subsection{Classic Threats}\label{sec:cThreats}
Classic threats to cryptographic systems may be categorized as follows (from weakest to most severe):  ciphertext-only attacks, known-plaintext attacks, and chosen-plaintext attacks.


In a ciphertext-only attack, the cryptanalyst might have some knowledge of the general system and the language used in the message, but the only significant data available to him is the encrypted transmission intercepted from the public channel.  In a known-plaintext attack, the cryptanalyst has knowledge of both the plaintext and its ciphertext counterpart.  To illustrate how a known-plaintext attack could take place, keep in mind that the rigid structure of most languages and business formalities often provides a cryptanalyst with a significant amount of a priori knowledge of the details of a possible plaintext message.  Armed with that knowledge and with an intercepted ciphertext message, the cryptanalyst can prepare a known-plaintext attack.

Suppose an encrypted message directs a foreign minister to make a particular public statement.  If the foreign minister makes the statement without any paraphrasing, a cryptanalyst may be privy to both the intercepted ciphertext and its plaintext translation for that encryption system.  Even though this type of attack is not always possible, a system is not considered to be secure unless it is designed to be secure against the plaintext attack \cite{sklar2001digital, DiffieHellman}.

In a chosen-plaintext attack, the cryptanalyst has the opportunity to select the plaintext which will then appear as ciphertext.  A chosen-plaintext attack was used by the United States during World War II to learn more about the Japanese Navy's intentions.  On May 20, 1942, Admiral Yamamoto, Commander-in-Chief of the Imperial Japanese Navy, issued an order describing detailed tactics that the Japanese Navy was to use when assaulting Midway Island.  This ciphertext message was not only intercepted by the Allied listening posts, by the time of its interception, the Americans had learned enough of the Japanese cryptographic system to be able to decrypt most of the message.  There were still some important parts, however, where the meaning of the message was uncertain to the Americans, such as the location of the assault.  The Americans suspected that the characters "AF" stood for Midway Island, but to be certain, Joseph Rochefort, head of the Combat Intelligence Unit, decided to use a chosen-plaintext attack to trick the Japanese into providing concrete evidence.  He had the Midway garrison broadcast a distinctive plaintext message in which Midway reported that its fresh-water distillation plant had broken down.  The American cryptanalysts only needed to wait approximately two days before they intercepted another Japanese encrypted message, which stated, "AF is short of fresh water" \cite{sklar2001digital, CodeBreakers}.

\section{Physical Layer Key Generation}\label{sec:physLayerKeyGen}
The purpose of physical layer key generation is ultimately to come up with a new practical way of generating secret keys between legitimate users who are using an insecure wireless channel, but without users having to endure the logistical and computational burdens associated with methods such as public key encryption.  Physical layer key generation specifically focuses on generating keys from the inherent randomness of wireless channels.  Instead of using pseudo-random number generators whose algorithms can be copied and results reproduced, physical layer key generation uses measurements of a channel that is varying with time, and which is not based on an algorithm and cannot be copied and reproduced, to generate random bits and form a secret key.  Instead of using public key cryptography to exchange secret keys over a wireless channel, physical layer key generation relies on the two legitimate users forming the same key based on their reciprocal channel measurements.  Physical layer key generation appears to be a promising alternative to public key cryptography in establishing secret keys between two users \cite{KeyGenWireChann}.

There are three basic principles for key generation that can be found in the literature \cite{KeyGenWireChann} and which we will briefly discuss.  Those principles are temporal variation, channel reciprocity, and spatial decorrelation.

Temporal variation occurs when there is movement in the channel, perhaps either from a mobile transmitter or receiver, or from any objects moving within the environment.  This unpredictable movement will cause changes in reflection and refraction, and a scattering of the channel paths.  It will create unpredictable randomness that can be used as the random source for generating secret keys.

Channel reciprocity indicates that any multipath or fading in the channel is identical at both ends of the same link and is the basis for the two end users being able to generate identical keys.  Since the signals ultimately have to be measured with hardware implementations that typically function in half duplex mode, the signals of the uplink and downlink paths are limited to time-division duplexing (TDD) systems and slow fading channels \cite{KeyGenWireChann}.  It has been experimentally demonstrated, while using the IEEE standard 802.11 which uses TDD, that channel reciprocity can be exploited \cite{ExpReciprocity, Reciprocity}.

Spatial decorrelation implies that an eavesdropper distanced more than one half-wavelength away from either legitimate user performing key generation will experience uncorrelated multipath fading and will not be able to generate the same key as either legitimate user.  We will claim the property of spatial decorrelation, since it is essential to ensure the security of the key generation process; however, we must also point out that spatial decorrelation and channel variation for illegitimate users may not be satisfied by all environments.  Considering Jake's model with a uniform scattering Rayleigh environment without a line-of-sight path, if the number of scatters increases to infinity, the signal will decorrelate over a distance of approximately one half-wavelength \cite{goldsmith2005wireless, jakes1974microwave}.  Jake's model is the type of environment we will assume.

\section{Problem Statement}\label{sec:problemStatement}
To avoid difficulties associated with secret key distribution and management, in recent years various physical layer approaches have been proposed for generating encryption keys using channel state information \cite{lai_etal_keygen2013}. These approaches use the inherent randomness of wireless channels and take advantage of the reciprocity properties of the channel between a wireless transmitter and its corresponding legitimate receiver to establish secret keys which may not be recreated by eavesdroppers overhearing the information exchanged over an uncorrelated channel.

Physical layer generation of secret keys avoids the need for key distribution or exchange, since the keys become known to both the transmitter and the legitimate receiver during the generation process.  Furthermore, with time-varying wireless channels, keys can be renewed dynamically using new measurements of the channel parameters.  For slowly varying 
channels, where the channel coherence time is large and the channel parameters change slowly in time, the rate of generating secret key elements becomes very small, and in such instances parasitic antenna arrays such as the RECAP \cite{RECAP, recap2012} or ESPAR \cite{espar2003} arrays may be used to randomize the channel and decrease channel coherence time.

Secret key generation based on noisy channel measurements at transmitter and receiver is discussed in \cite{LDPC_code_const,LDPC_Quant}, where the use of Slepian-Wolf coding \cite{sun_etal_wcnc2010,Secrecy_Joint_Gaussian_RV,Key_Gen_Fading,Info_Theo_Security,Key_Gen_Diversity} with LDPC codes for key reconciliation is studied.  We note that channel-based key generation approaches use knowledge of the statistical properties of the channel, which requires extensive time-consuming measurements to estimate channel statistics.

In this thesis, we study practical methods for secret key generation that have low complexity and are based on one-bit scalar quantization of the real and imaginary components of the complex channel gain between the transmitter and the legitimate receiver. The proposed method exploits the symmetry of the probability density function (PDF) of the channel gain and  requires no prior knowledge of the channel variance. Thus, no channel observation and variance estimation is needed, which speeds up the key generation. Furthermore, the method simplifies the logarithmic-likelihood ratio (LLR) calculation used in the key generation for fast computation.  Specifically, we propose a novel way of evaluating LLRs that is based on the difference between the channel estimates at the legitimate nodes, while it combines advantage distillation with information reconciliation and privacy amplification~\cite{LDPC_code_const, LDPC_Quant} for key reconciliation using LDPC codes.

Our approach involves using the following:

\begin{itemize}
\item Low complexity, one-bit scalar quantization
\item The real and imaginary components of the complex channel gain
\item The symmetry of the probability density function of the channel gain
\item No prior knowledge of the channel variance
\item No channel observation
\item No variance estimation
\item Novel simplified LLR calculations
\end{itemize}

\newpage

\section{Thesis Outline}\label{sec:thesisOutline}
The thesis is organized in the following chapters:  in Chapter~\ref{sec:agreement}, we introduce the system model and we explain the detailed concepts of our proposed key generation mechanism.  In Chapter~\ref{sec:propLLRs}, we present the channel quantization scheme used by our proposed key generation approach and we give details of the LLR calculations.  In Chapter~\ref{sec:LDPC}, we explain details concerning the Low-Density Parity-Check codes that are pertinent to our secret key generation scheme and simulation results.  In Chapter~\ref{sec:sims}, we present numerical results obtained from simulations and discuss the performance of the proposed method, comparing it with the "censoring" scheme for key generation that is widely used in the literature \cite{SecretKeyReactance,Cens_mil}, and we present the proposed LLR calculations, comparing them to approximate LLR calculations.  We conclude the thesis with final remarks in Chapter~\ref{sec:con}.

\chapter{Secret Key Agreement}\label{sec:agreement}
This chapter discusses a wireless security design for secret key agreement at the physical layer for mobile communication networks.  The three phases of this wireless security design are introduced (Advantage Distillation, Information Reconciliation, and Privacy Amplification) and a high-level understanding of how these three phases of wireless security operate is discussed.  Before these three phases of wireless security are presented, a review of several wireless communications terms is offered:

\begin{enumerate}
\item Time-division duplex (TDD) is a scheme in which duplexing is achieved by separate transmit and receive time intervals rather than by separate frequencies \cite{introBook}.  The uplink is separated from the downlink by the allocation of different time slots in the same frequency band and the transmit and receive functions are toggled over a given time interval.  TDD allows both nodes of a communication link to transmit to one another using the same frequency.
\item Coherence time $t_c$ is the time duration over which path loss remains approximately constant and the time over which the channel impulse response is considered to be not varying \cite{introBook}.  Coherence time $t_c$ is the time domain dual of the Doppler spread and is used to characterize the time-varying nature of the frequency dispersiveness of the channel in the time domain \cite{introBook, NIwhitepaper}.  Coherence time is a statistical measure of the time duration over which the channel impulse response is essentially invariant and quantifies the similarity of the channel response at different times \cite{NIwhitepaper}.  In other words, coherence time is the time duration over which two received signals have a strong potential for correlation.  The definition of coherence time implies that two signals arriving with a time separation greater than $t_c$ are affected differently by the channel \cite{NIwhitepaper}.
\end{enumerate}


\section{Advantage Distillation}\label{sec:distillation}
\begin{figure}
\centering
\includegraphics[scale=1]{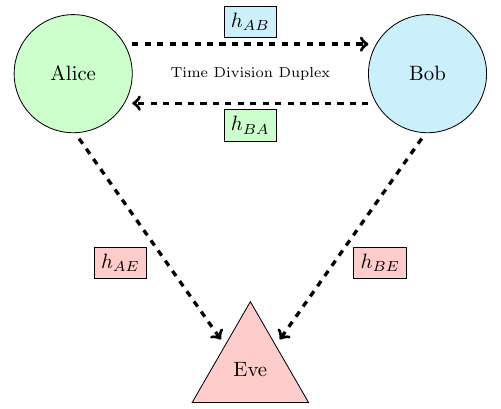}
\caption[Advantage Distillation.]{\textbf{Advantage Distillation.}}\label{fig:sys_mod}
\end{figure}

Figure~\ref{fig:sys_mod} depicts two nodes, Alice and Bob, of a mobile wireless network actively communicating with one another while a potential eavesdropper, Eve, listens in on the communication.  Specifically, Alice and Bob are performing advantage distillation to begin forming a new secret key so that their communications can be encrypted and therefore be indiscernible to Eve.  In Figure~\ref{fig:sys_mod}, Alice and Bob take advantage of their unique fading channel path and take measurements of their fading channel path using a pilot sequence.  Alice and Bob now know more about their shared fading channel path's noise characteristics than the eavesdropper, Eve, and can begin the process of agreeing upon a secret key over an insecure channel.

This advantage distillation process consists of a TDD channel between Alice and Bob and a pilot sequence that is known publicly.  Alice and Bob exchange the pilot sequence across the channel in TDD mode.  When Alice and Bob detect the pilot sequence at their receivers, they eliminate the pilot sequence in order to obtain a measure of the channel characteristics, termed channel state information (CSI).

As long as the receiving antennas of Alice, Bob, and Eve are physically separated by more than half a wavelength, $\lambda$, from each other, the different channels (Alice and Bob, Alice and Eve, Bob and Eve) will be independent of one another and will experience independent channel characteristics.  This independent CSI enables Alice and Bob to gain an advantage over Eve through the distillation of the random noise characteristics experienced by the different channels.  Alice and Bob will ultimately generate a secret key based on the independent random CSI obtained from exchanging pilot sequences.  The advantage distillation, as performed in this specific scenario, will be referred to as secret key generation interchangeably throughout the remainder of this chapter.

\begin{figure}
\centering
\includegraphics[scale=1]{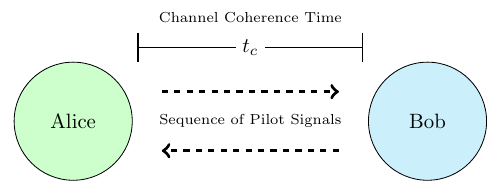}
\caption[A Single Channel Measurement.]{\textbf{A Single Channel Measurement.}}\label{fig:seqPS}
\end{figure}

Looking more closely at the details of this secret key generation, Alice and Bob generate a secret key based on measurements of their wireless channels, which are assumed to be randomly changing over time with a known channel coherence time $t_c$.  Alice and Bob measure the channel during each channel coherence time by transmitting a sequence of pilot signals to each other in TDD.  The assumption is made that the channel does not change during the transmission of one sequence of pilot signals.  In other words, looking at Figure~\ref{fig:seqPS} as a visual aid, the time required to transmit one sequence of pilot signals is less than the channel coherence time $t_c$.  Alice and Bob wait until the next channel coherence time before re-exchanging the sequence of pilot signals and measuring the next channel instance.  These timing assumptions are made to ensure that the following conditions exist:

\begin{enumerate}
\item Each time Alice and Bob measure the channel (i.e., once every $t_c$), both Alice and Bob will be measuring the same channel instance and their CSI will be almost identical, allowing them to generate identical secret keys
\item Each measurement of the channel will be uncorrelated to the previous measurements of the channel to allow inherent randomness to exist in the generated secret key
\end{enumerate}

Another way of thinking about Figure~\ref{fig:seqPS} is that each dash in the dashed lines labeled "Sequence of Pilot Signals" represents one pilot signal within the sequence.  One sequence of pilot signals must occur within the channel coherence time to ensure that Alice and Bob are measuring the same channel instance and are obtaining the same measurements.  Each consecutive channel measurement must occur over a different channel instance to ensure that the generated secret key will contain inherent randomness provided by the channel.

Upon transmission of the pilot signals by Alice and Bob, the corresponding channel estimates at time instant $n$ for Alice and Bob are, respectively:

\begin{align}
a[n] = h_{BA}[n] + w_a[n] \label{eq:alice_sample}\\
b[n] = h_{AB}[n] + w_b[n] \label{eq:bob_sample}
\end{align}

\hfill \break
where $h_{BA}[n]$ and $h_{AB}[n]$ are zero mean circular complex Gaussian random variables, and $w_a[n]$ and $w_b[n]$ denote the zero mean additive white Gaussian noise corrupting the channel measurements at Alice and Bob with variances $\sigma^2_a$ and $\sigma^2_b$, respectively.  For simplicity, it is assumed that noise variances are equal, $\sigma^2_a = \sigma^2_b = \sigma^2_w$, and that, due to TDD mode, consecutive channel measurements done by both Alice and Bob in $t_c$ period of time are identical, i.e., $h_{BA}=h_{AB}$.

We assume that Eve has infinite resources available and knows the pilot signals, so she can measure her
corresponding channels to Alice and Bob $h_{AE}$ and $h_{BE}$, respectively, using Alice and Bob's
pilot transmissions.  However, we assume that Eve is several wavelengths away from both Alice and Bob
so that the channels $h_{AE}$, $h_{BE}$, and $h_{BA}$ are all uncorrelated and Eve is not able to extract meaningful information related to Alice and Bob's measurements or generated secret key.  Furthermore, Eve is assumed to be a passive eavesdropper, one who always listens but never transmits.

In practical scenarios, Alice and Bob are not guaranteed to have measured the exact same CSI.  Therefore, information reconciliation and then privacy amplification must be performed to ensure that Alice and Bob agree upon the same key and to ensure that Eve cannot obtain the same secret key.



\section{Information Reconciliation}\label{sec:infoRec}
Once Alice and Bob obtain CSI by performing advantage distillation, Alice and Bob are not guaranteed to have identical CSI due to independent noise effects and due to independent quantization effects that occur at the separate ends of the reciprocal link.  Therefore, Alice and Bob must reconcile the differences between their CSI in order to agree upon the same secret key.

In order to reconcile the CSI over an insecure channel without giving away the secret key, concepts from both source coding and channel coding are utilized.  Slepian-Wolf coding with side information is employed from source coding, and Low-Density Parity-Check (LDPC) codes are employed from channel coding.

\begin{figure}
\centering
\includegraphics[scale=1]{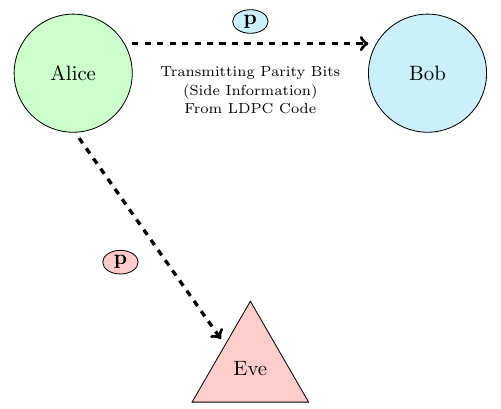}
\caption[Information Reconciliation.]{\textbf{Information Reconciliation.}}\label{fig:infoRec}
\end{figure}

Reference \cite{sun_etal_wcnc2010} introduces and goes into detail about employing Slepian-Wolf coding with side information by using capacity approaching LDPC codes while References \cite{LDPC_code_const} and \cite{LDPC_Quant} further investigate the use of LDPC codes to employ Slepian-Wolf coding for information reconciliation.  To see and understand how information reconciliation is performed over the insecure channel between Alice and Bob, we will explain the situation displayed in Figure~\ref{fig:infoRec}.

Figure~\ref{fig:infoRec} depicts Alice and Bob performing information reconciliation after advantage distillation so that their CSI can agree and can be used to form a secret key.  This information reconciliation consists of Alice forming an LDPC code from her CSI and transmitting only the parity bits to Bob.  In order to ensure that the parity bits are received by Bob correctly, they can be encoded using another LDPC code and transmitted by modulating a carrier signal with binary phase-shift keying (BPSK) \cite{LDPC_Quant}.

Once Bob receives the parity bits from Alice's CSI, Bob can decode his CSI with Alice's parity bits and can reconcile his CSI to Alice's CSI by using the LDPC decoder.  LDPC codes commonly implement soft-decision decoding using the sum-product algorithm.  To make soft-decisions, the decoder cannot just be handed bits of information, but rather must be given bits of information along with the likelihood ratios that those bits are correct.  In many cases where likelihood ratios are required, the logarithmic-likelihood ratio (LLR), or the log of the likelihood ratio, can actually be less computationally expensive and is therefore used.  Reference \cite{LDPC_Quant} proposes one method for calculating LLRs for information reconciliation; however, the proposed method is quite computationally expensive.

In the context of channel coding, the LLR is calculated by determining the probability that symbol $x$ was transmitted, given that symbol $x$ was received.  In the context of information reconciliation, however, we must determine the probability that Alice quantized her CSI for a given symbol to quantization region $x$, given that Bob quantized his CSI for that given symbol to quantization region $x$.

Developing the LLRs for LDPC codes is an integral part of information reconciliation, and our proposed LLR calculation for information reconciliation is discussed and examined in detail in Chapter~\ref{sec:propLLRs}.  By using the LLRs developed for the specific advantage distillation used to characterize the CSI, Bob is able to resolve the discrepancies between his and Alice's CSI so that his CSI matches Alice's CSI and they agree upon a secret key.  Of course, Eve will attempt to perform information reconciliation using Alice's parity bits as well, in order to crack Alice and Bob's secret key, but without the correlated channel measurements that Alice and Bob have, Eve will be unsuccessful.

\begin{figure}
\centering
\includegraphics[scale=0.85]{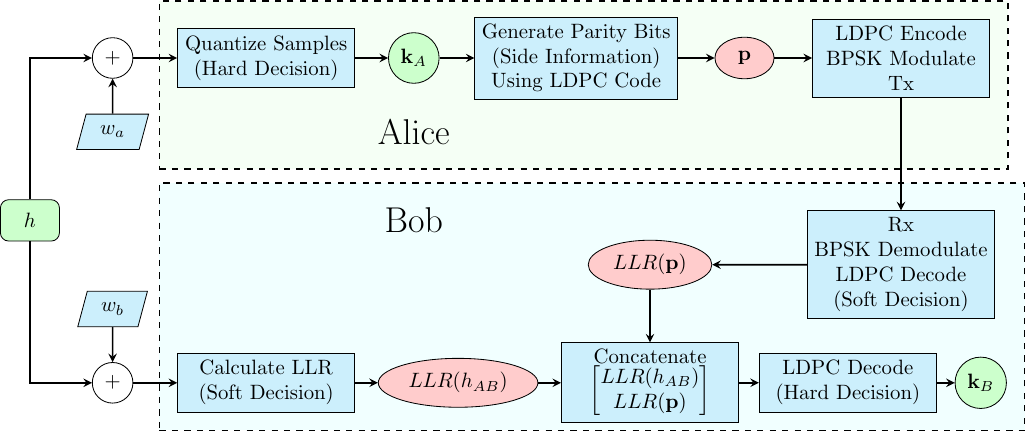}
\caption[Proposed Key Generation Mechanism.]{\textbf{Proposed Key Generation Mechanism.}}\label{fig:sys_block_diag}
\end{figure}

In Figure~\ref{fig:sys_block_diag}, we display a more detailed flowchart of our proposed key generation mechanism.  When Alice receives a pilot signal from Bob, Alice quantizes her channel measurements making hard-decisions (for example, Alice may either quantize her channel measurement to a $1$ bit or to a $0$ bit).  However, when Bob receives a pilot signal from Alice, Bob will not quantize his channel measurements making hard-decisions.  Bob will calculate LLRs of his channel measurements making soft-decisions (for example, Bob may end up with a $-99$ or a $+105$, where the negative symbol corresponds to a $1$ bit, the positive symbol corresponds to a $0$ bit, and the magnitude corresponds to how likely it is that Alice quantized her channel measurement to the same quantization region).  The greater the magnitude of the soft-decision, the more likely Alice quantized her channel measurement to the same quantization region as Bob.

Next, Alice forms her secret key based on her quantized samples and then generates parity bits, or side information, about her secret key using an LDPC code.  This side information (and only the side information, not the secret key with it) is then transmitted to Bob using BPSK modulation with additional forward error correction to ensure that Bob receives the correct parity bits from Alice.  In other words, Alice generates the side information, encodes it with a separate LDPC code for forward error correction, and transmits it over to Bob using BPSK modulation.

When Bob receives the transmitted signal from Alice's BPSK transmission, Bob performs BPSK demodulation and then performs LDPC decoding, all the while only making soft-decisions.  After Bob is finished making soft-decisions, he has the LLRs from his channel measurements, and he also has the LLRs from the BPSK transmission from Alice.  Bob then concatenates the LLRs of Alice's parity bits to the end of his channel measurement LLRs in order to put them through the LDPC decoder, to form hard-decisions, and thereby to obtain his secret key.

To summarize this setup, we take advantage of the correlation between channel measurements at Alice and Bob and we employ Slepian-Wolf coding for reconciliation of the secret key bits generated by Alice and Bob \cite{LDPC_code_const}.  This involves partial information exchange between Alice and Bob, which consists of parity bits (side information) generated at Alice by using LDPC codes.  The side information is then encapsulated by a second LDPC code block for forward error correction and sent to Bob over a public channel, as shown schematically in Figure~\ref{fig:sys_block_diag}.  Thus, our proposed approach for key generation involves two independent LDPC code blocks, one for generating the side information to employ Slepian-Wolf coding, and the other one for protecting the side information from errors during transmission over the public channel.  We note that, even if Eve is able to intercept the side information transmitted over the public channel, she will not be able to reconstruct the secret key bits generated by both Alice and Bob because she will not have the additional information needed, which depends on the $h_{BA}$ channel measurements.

\section{Privacy Amplification}\label{sec:privAmp}
Privacy amplification involves taking steps to enhance the privacy of a secret key.  One of the advantages that LDPC codes (along with other soft-decision coding algorithms) provide when they are employed to complete the information reconciliation process is a phenomenon that provides a disadvantage to the channel coding (or error correction) process.  Without enough correct information to start with, the sum-product decoding algorithms that are implemented in LDPC decoders will never be able to converge on the correct result \cite{MITOpenCourseWare}.  This fact is what prevents Eve from being able to succeed when performing the information reconciliation process; however, the fact still remains that Eve has some knowledge of the secret key.  Therefore, an additional step, termed privacy amplification, must still be performed by Alice and Bob in order to protect their secret key from both ciphertext-only and known-plaintext attacks.

\begin{figure}
\centering
\includegraphics[scale=0.75]{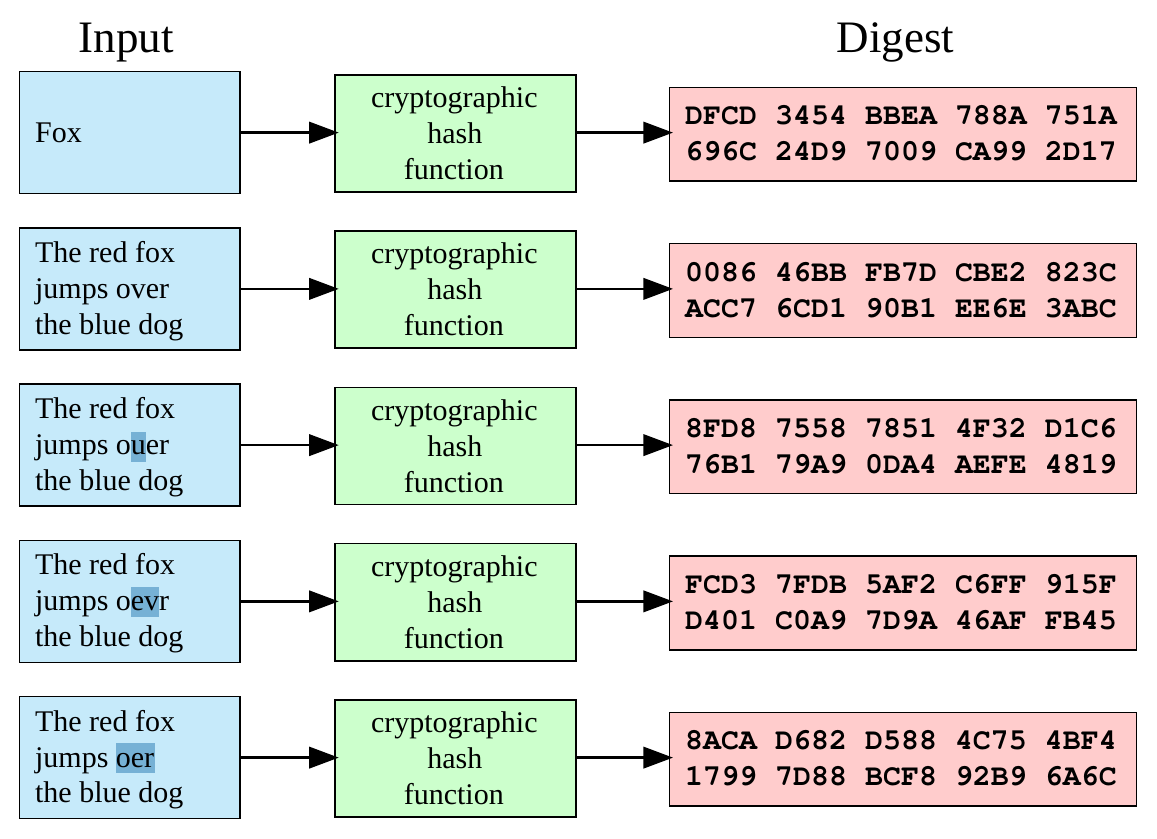}
\caption[The Cryptographic Hash Function SHA-1.]{\textbf{The Cryptographic Hash Function SHA-1.}  This figure is recreated from the Public Domain.}\label{fig:hash}
\end{figure}

To significantly decrease the probability that Eve can obtain the same key as Alice and Bob, privacy amplification must be performed after information reconciliation.  Privacy amplification can be performed through the use of an effective cryptographic hash function, an example of which is shown in Figure~\ref{fig:hash}.  An effective cryptographic hash function should generate digests such that small changes (even single bit differences) in the input will create completely different digests at the output.  SHA-1, demonstrated in Figure~\ref{fig:hash}, is an example of an effective cryptographic hash function.


A cryptographic hash function is the final stage of secret key agreement that Alice and Bob go through to generate their secret key.  The cryptographic hash function can actually be agreed upon publicly, so that Eve will also have knowledge of the cryptographic hash function.  However, as long as an effective cryptographic hash function is utilized (such that small changes in the input create completely different digests at the output), Eve will have to perfectly reconcile her intercepted channel measurement information to Alice's CSI in order to be able to crack Alice and Bob's new secret key.  As long as Eve is more than half a wavelength, $\lambda$, away from Alice and Bob, the probability of Eve successfully performing information reconciliation is close to zero.

By using the output (or the digest) of the cryptographic hash function as the secret key, Alice and Bob can prevent Eve from being able to use the information that she has already obtained about the secret key to further attempt to crack the secret key with ciphertext-only attacks or known-plaintext attacks.  Even though Eve may still attempt ciphertext-only attacks and known-plaintext attacks, the use of the cryptographic hash function renders the parity bit information and Eve's channel measurements worthless unless Eve can completely reconcile her CSI to Alice's CSI.  Essentially, Eve is left to use brute forcing methods to crack Alice and Bob's new secret key.

By designing the output (or the digest) of the cryptographic hash function to be a sufficient number of bits in length (e.g., if the digest creates a 4096 bit key), then even if Eve were to attempt to brute-force her way to obtaining the same key as Alice and Bob, Alice and Bob could generate a new secret key before Eve successfully cracked their current secret key.  If, for example, Alice and Bob successfully generated a 4096 bit key and Eve was left to use brute forcing methods to crack the key, there would be $2^{4096}\approx10^{1233}$ possible keys.  In an exhaustive search, one might expect to crack the key approximately halfway through the search.  If we assumed that each trial time required a computation time of 0.001 ns and that 100 billion computations could be performed in parallel, then the completed search time would require more than $1.5\times10^{1202}$  years!  This would allow enough time for Alice and Bob to generate a new secret key before the previous one was cracked by Eve.

By performing advantage distillation, information reconciliation, and privacy amplification in a mobile environment as described above, Alice and Bob now have the ability to agree upon a secret key with unconditional secrecy over an insecure wireless channel.

\chapter{Log-Likelihood Ratios (LLRs)}\label{sec:propLLRs}
In order for physical layer secret key agreement to work effectively and practically, the process of secret key generation and information reconciliation needs to succeed consistently and reliably.  To increase the success rate for physical layer secret key agreement, different secret key generation schemes have been proposed, LLR calculation methods have been proposed, and new strategies for designing the parity bit structure of information reconciliation codes have even been proposed.

When a new key generation scheme is proposed, typically the new approach centers around improving Alice and Bob's ability to independently measure their reciprocal channels to obtain highly or completely correlated results.  However, at the same time, the new approach cannot significantly increase the ability of Eve to measure, or the probability that Eve can measure, results that are correlated to Alice and Bob's CSI.

One proposed method for improving secret key generation, which is now widely seen in the literature \cite{SecretKeyReactance, Cens_mil}, is termed the "censoring" scheme and is based on a censoring approach.  Specifically, in order to reduce the number of bit disagreements between the generated keys, in \cite{Cens_mil} the channel measurement samples that fall in the region $[-\gamma,\gamma]$ at Alice's side are eliminated (censored) and they are not used for key generation, since bit disagreements between the generated keys are more likely to occur at those samples.  The indices of the censored symbols are sent to Bob.  The expected number of censored samples can be adjusted by changing the $\gamma$ value.  As $\gamma$ increases, more samples are eliminated, leading to a lower probability of bit disagreement between keys as well as a slower key generation (in terms of bits generated per second).  For a channel with a short $t_c$ coherence time, a fixed number of bits for a key can be generated quickly; therefore, censoring some of the channel measurements is a meaningful approach to decrease the probability of bit disagreements between keys.  The main disadvantage of the censoring scheme is that Alice needs to transmit the indices of the samples that are censored or discarded in the key generation to Bob, since not all of the samples are used for key generation.

In the literature, it is not uncommon to see secret key generation discussed as though it is completely separate and isolated from the information reconciliation process.  However, as has been presented in Chapter~\ref{sec:agreement}, we note that, in order to improve the probability of successfully correcting bit disagreements between Alice and Bob, Bob must calculate LLRs (or develop soft-decisions initially instead of hard-decisions) to aid in the information reconciliation process.  These LLR calculations, in order to work optimally, must be determined based both on the channel measurement quantization scheme and on the likelihood that Alice quantized a given channel measurement to the same quantization region as Bob. 

In \cite{LDPC_Quant}, an approach to tie both the secret key generation procedure and the information reconciliation procedure together was presented.  This technique began with complex LLR calculations, which led to unique channel quantization schemes designed to reduce the complexity of the LLR calculations.  Adding to the complexity of the LLR calculations, this approach requires knowledge of the statistical properties of the channel, which requires extensive time-consuming measurements to estimate channel statistics.  In particular, prior knowledge of the channel variance is necessary, in order to perform the LLR calculations presented in \cite{LDPC_Quant}.

In this thesis, a practical method for combining key generation with information reconciliation is presented.  It has low complexity and is comparable to the "censoring" scheme discussed earlier.  The secret key generation is based on one-bit scalar quantization of the real and imaginary components of the complex channel gain between the transmitter and the legitimate receiver.  The proposed method exploits the symmetry of the probability density function of the channel gain and requires no prior knowledge of the channel variance.  Thus, no channel observation and variance estimation are needed, which speeds up the key generation.  Furthermore, the method simplifies the LLR calculation used in the key generation for fast computation.  Specifically, we propose a novel way of evaluating LLRs that is based on the difference between the channel estimates at the legitimate nodes.  This approach allows a low complexity key generation scheme, as well as the "censoring" scheme, to be combined with key reconciliation using LDPC codes.

The secret key bits are generated at Alice and Bob simultaneously by exploiting the random characteristics of the wireless channel between them. Specifically, both Alice and Bob measure their corresponding (reciprocal) channels until the required number of samples for a secret key are obtained, and then they generate the key based on the vector of sampled channel measurements. However, due to the independent noise corrupting Alice and Bob's channel measurements, bit disagreements between their generated keys might occur.  The probability of bit disagreement will be evaluated in Chapter~\ref{sec:sims} using Monte-Carlo simulations.

In the proposed key generation scheme, we assume that Alice is the leader node who quantizes her channel measurements by making hard-decisions, and thereby generating her secret key bits $\kv_A$, while Bob is the follower who computes his secret key bits $\kv_B$ by combining LLRs of his channel measurements, along with LLRs of the side information from Alice. Furthermore, Alice uses both the real and imaginary parts of the channel samples independently to generate independent key bits by 1-bit scalar quantization.  Specifically, Alice quantizes the obtained samples with the following function to generate key bits: 

\begin{align}
k_A[n] = 
     \begin{cases}
       0, &\quad e[n] > 0\\
       1, &\quad e[n] \leq 0
     \end{cases}
\end{align}

\hfill \break
where $e[n] \in \{ \operatorname{Re}(a[n]) , \operatorname{Im}(a[n]) \}$ and $a[n]$ is defined in equation~(\ref{eq:alice_sample}).
This is followed by generation of the parity vector $\pv$ by Alice using the first LDPC encoder matrix $H_p$ for Slepian-Wolf coding.  This parity vector $\pv$ is then encapsulated with the second LDPC encoder matrix $H_t$.  The second LDPC encoder is simply to allow the parity vector $\pv$ to be transmitted to Bob with traditional forward error correction and binary phase-shift keying (BPSK) techniques.


At the other side, Bob measures his channel from Alice's pilot signals, but instead of forming hard-decisions as Alice does, Bob calculates the LLRs for Alice's measurements based on his own observations.  Additionally, Bob receives the side information $\pv$ from Alice's BPSK transmission and he demodulates the BPSK signal, calculating conventional LLRs.  Bob extracts the parity vector $\pv$ using, once again, soft-decision decoding, as opposed to hard-decision decoding.  Bob now has LLRs for Alice's channel measurements, along with LLRs for Alice's parity vector $\pv$.  Note that the channel samples and the received BPSK signals have different random properties, thus they require different LLR calculations, as will be described in detail.  In the end, Bob concatenates both his channel measurement LLRs and side information LLRs and applies Slepian-Wolf decoding  using the $H_p$ matrix to determine his secret key $\kv_B$, which is correlated with $\kv_A$.  We note that, ideally, $\kv_A = \kv_B$, although some bit disagreements between $\kv_A$ and $\kv_B$ are possible, as discussed in Chapter~\ref{sec:sims}.  The two types of LLR calculations used in our approach, shown in Figure~\ref{fig:sys_block_diag}, will be discussed in detail in the following sections.

\section{LLR Calculations for Side Information}\label{sec:LLRCalcSideInfo}
When Alice transmits her side information to Bob, she uses traditional communication methods.  In our case, we use a simple BPSK modulation technique.  In order for Bob to optimize his key reconciliation, he does not form hard-decisions of Alice's parity bits at the output of the BPSK demodulator; instead, he will calculate LLRs based on conventional BPSK demodulation techniques.  Figure~\ref{fig:bpsk} illustrates the conditional PDFs for the binary noise-perturbed output signals, $b[n] = a_0 + n_0$ and $b[n] = a_1 + n_0$ from a typical receiver.  The signals $a_0$ and $a_1$, represented by their bit-energy levels $\sqrt{E}$ and $-\sqrt{E}$ in Figure~\ref{fig:bpsk}, are mutually independent and are equally likely.  The noise $n_0$ is assumed to be an independent Gaussian random variable with zero mean, variance $\sigma^2_w$, and PDF given by

\begin{align}
p(n_0) = \frac{1}{\sqrt{2\pi\sigma^2_w}} e^{\frac{-n^2_0}{2\sigma^2_w}}.  \label{eq:noise}
\end{align}

\hfill \break

Since the constellation points for the BPSK demodulator have coordinates $(\sqrt{E},0)$ and $(-\sqrt{E},0)$, the LLRs can be calculated as follows:

\begin{figure}
\centering
\includegraphics[scale=1]{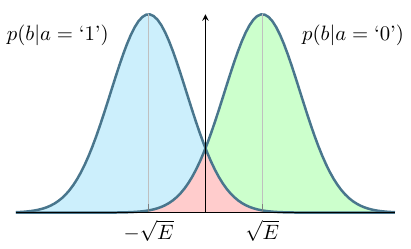}
\caption[Conditional PDFs for a Typical Binary Receiver.]{\textbf{Conditional PDFs for a Typical Binary Receiver.}}\label{fig:bpsk}
\end{figure}

\begin{align}
LLR_{side}[n] &= \ln\frac{\mbox{Pr}(a[n]=\text{`0'}|b[n])}{\mbox{Pr}(a[n]=\text{`1'}|b[n])} \label{eq:approxLLR1} \\
LLR_{side}[n] &= \ln\frac{p(b[n]|a[n]=\text{`0'})\frac{\mbox{Pr}(a[n]=\text{`0'})}{p(b[n])}}{p(b[n]|a[n]=\text{`1'})\frac{\mbox{Pr}(a[n]=\text{`1'})}{p(b[n])}} \label{eq:approxLLR2} \\
LLR_{side}[n] &= \ln\frac{p(b[n]|a[n]=\text{`0'})}{p(b[n]|a[n]=\text{`1'})}. \label{eq:approxLLR3}
\end{align}

\hfill \break
Equation~(\ref{eq:approxLLR1}) mathematically represents the idea of determining the ratio of the conditional probability that Alice transmitted a `0' bit at sample $n$ given that Bob received value $b$ at sample $n$ to the conditional probability that Alice transmitted a `1' bit at sample $n$ given that Bob received value $b$ at sample $n$.  Exploiting Bayes' Theorem, equation~(\ref{eq:approxLLR1}) transitions to equation~(\ref{eq:approxLLR2}).  Since the probability that Alice transmitted a `0' bit at sample $n$ is equal to the probability that Alice transmitted a `1' bit at sample $n$, the quantities $\mbox{Pr}(a[n]=\text{`0'})$ and $\mbox{Pr}(a[n]=\text{`1'})$ cancel out and equation~(\ref{eq:approxLLR2}) reduces to equation~(\ref{eq:approxLLR3}).  We can now plug in the PDF of the noise distribution, see equation~(\ref{eq:noise}), into equation~(\ref{eq:approxLLR3}) and obtain equation~(\ref{eq:approxLLR4}).  Further simplifying equation~(\ref{eq:approxLLR4}), we can ultimately reduce the LLR calculation to equation~(\ref{eq:approxLLR6}):

\begin{align}
LLR_{side}[n] &= \ln\frac{ \frac{1}{\sqrt{2\pi\sigma^2_w} } e^{ \frac{-(b[n]-\sqrt{E})^2}{2\sigma^2_w}} }{ \frac{1}{\sqrt{2\pi\sigma^2_w} } e^{ \frac{-(b[n]+\sqrt{E})^2}{2\sigma^2_w}} } \label{eq:approxLLR4} \\
 &= \ln\frac{ e^{ \frac{-(b[n]-\sqrt{E})^2}{2\sigma^2_w}} }{ e^{ \frac{-(b[n]+\sqrt{E})^2}{2\sigma^2_w}} } \nonumber \\
 &= \ln e^{ \frac{-(b[n]-\sqrt{E})^2}{2\sigma^2_w} - \frac{-(b[n]+\sqrt{E})^2}{2\sigma^2_w}} \nonumber \\
 &= \frac{-(b[n]-\sqrt{E})^2}{2\sigma^2_w} - \frac{-(b[n]+\sqrt{E})^2}{2\sigma^2_w} \nonumber \\
 &= \frac{-(b[n]-\sqrt{E})^2 + (b[n]+\sqrt{E})^2}{2\sigma^2_w} \nonumber \\
 &= \frac{(b[n]+\sqrt{E})^2-(b[n]-\sqrt{E})^2}{2\sigma^2_w} \nonumber \\ 
 &= \frac{(b^2[n]+2b[n]\sqrt{E}+E)-(b^2[n]-2b[n]\sqrt{E}+E)}{2\sigma^2_w} \nonumber \\
 &= \frac{2b[n]\sqrt{E}+2b[n]\sqrt{E}}{2\sigma^2_w} \nonumber \\
 &= \frac{4b[n]\sqrt{E}}{2\sigma^2_w} \nonumber \\
LLR_{side}[n] &= \frac{2b[n]\sqrt{E}}{\sigma^2_w}. \label{eq:approxLLR6}
\end{align}

\hfill \break

In order to perform BPSK transmission and reception in our simulations efficiently and reliably, we employed the BPSK modulator and demodulator objects in MATLAB\textsuperscript{\textregistered}'s Communications System Toolbox\textsuperscript{TM} \cite{MATLAB:2016}.  In the BPSK demodulation software, LLRs can be calculated using one of two approaches:  either the "logarithmic likelihood ratio" approach can be specified, which calculates equation~(\ref{eq:approxLLR4}); or, the "approximate logarithmic likelihood ratio" approach can be specified, which calculates equation~(\ref{eq:approxLLR6}) \cite{MATLAB:2016, viterbi}.

Even though both equations~(\ref{eq:approxLLR4})~and~(\ref{eq:approxLLR6}) theoretically produce the same result, MATLAB\textsuperscript{\textregistered} explains that "if the noise variance is very small (i.e., SNR is very high), log-likelihood ratio (LLR) computations can yield $\infty$ or $-\infty$.  This variance occurs because the LLR algorithm computes the exponential of very large or very small numbers using finite precision arithmetic.  As a best practice in such cases, use approximate LLR because this option's algorithm does not compute exponentials"~\cite{MATLAB:2016}.

\begin{figure}
\centering
\includegraphics[scale=0.5]{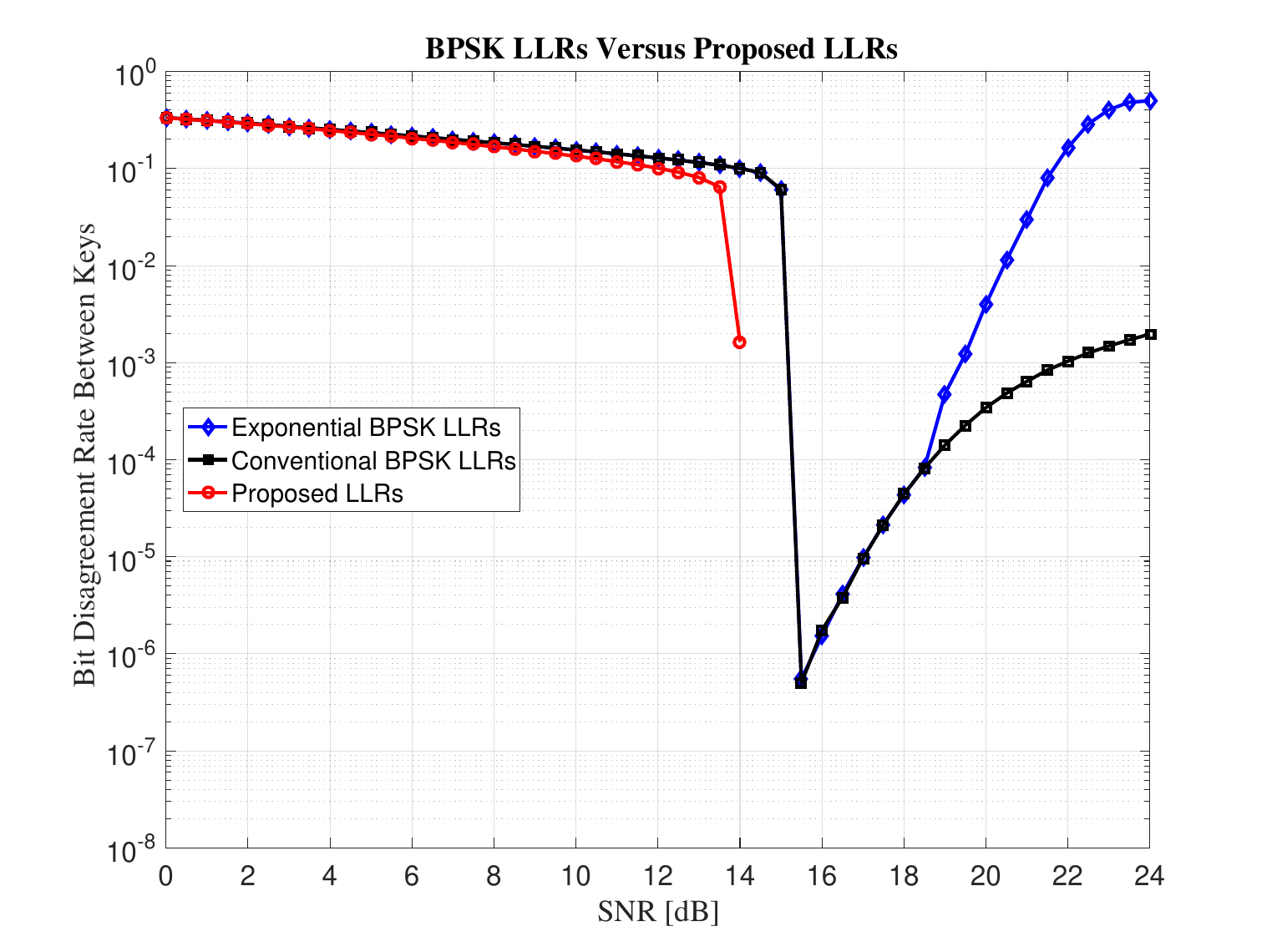}
\caption[BPSK LLRs Versus Proposed LLRs.]{\textbf{BPSK LLRs Versus Proposed LLRs.}}\label{fig:compareLLRs}
\end{figure}

To see an example of how the LLR computations yielding $\infty$ or $-\infty$ can detrimentally affect the bit error rate (BER) of a received signal, observe the diamond curve (exponential BPSK LLRs) in Figure~\ref{fig:compareLLRs}.  Although the diamond curve in Figure~\ref{fig:compareLLRs} is not just a BPSK transmission scheme on its own, BPSK transmission was involved, and demodulation using equation~(\ref{eq:approxLLR4}), the "logarithmic likelihood ratio" option in MATLAB\textsuperscript{\textregistered}, was employed to generate soft-decision outputs.  As can be seen, once the SNR increases above approximately 18~dB, the bit disagreement rate should continue to follow the square curve (conventional BPSK LLRs); however, due to the LLR algorithm computing exponentials of very large and very small numbers (at high SNR) using finite precision arithmetic, the bit disagreement rate increased to a worse level than it was at 0~dB!  The soft-decision outputs of the demodulator yielded $\infty$ or $-\infty$ at high SNR, which introduced significant error in the final stage of decoding.


In addition to reduced computational complexity, the diamond curve and the square curve in Figure~\ref{fig:compareLLRs} exemplify why we chose to use equation~(\ref{eq:approxLLR6}), the square curve, or the "approximate logarithmic likelihood ratio" option in MATLAB\textsuperscript{\textregistered}, when calculating soft-decision outputs from a BPSK demodulator.  The errors generated at high SNR values when using equation~(\ref{eq:approxLLR4}), the diamond curve, to calculate soft-decision outputs defeat the purpose of forming soft-decisions in the first place.

\section{LLR Calculations for Channel Measurements}\label{sec:LLRCalcChannel}
Having gone through the equations to calculate conventional BPSK LLRs, we will now look at the equations to calculate LLRs for the channel.  Since our channel quantization method only uses one bit, an initial idea may be simply to try using the conventional BPSK LLRs for Bob's channel measurements to see how well they work.  This approach is one that we applied for a benchmark and, not surprisingly, it worked to an extent.  The most detrimental effect that conventional BPSK LLRs have on channel measurements occurs when the SNR is relatively large.  Observing the square curve in Figure~\ref{fig:compareLLRs}, it can be seen that when conventional BPSK LLRs are used to generate LLRs for Bob's channel measurements, the approximation no longer holds valid, once the SNR increases beyond about 15~dB.  The BPSK approximation makes Bob's channel measurements overconfident and they can no longer be properly corrected by information reconciliation.  This effect appears similar to the effect of performing exponential LLR calculations using finite precision arithmetic at high SNR, as seen in the diamond curve of Figure~\ref{fig:compareLLRs}, although the error that is generated in this case, as seen in the square curve of Figure~\ref{fig:compareLLRs}, is not quite as drastic.


In order to improve upon the information reconciliation process, we must consider what information is being provided by Bob's channel measurements and we must consider how to effectively capture that information.  The BPSK LLRs capture information based on the assumptions (or questions) "What is the likelihood that Bob will receive a `0' given that Alice transmitted a `0'?" and "What is the likelihood that Bob will receive a `0' given that Alice transmitted a `1'?"  The assumptions (or questions) that we should be trying to capture are "What is the likelihood that Alice quantized her channel measurement to a `0', given Bob's channel measurement?" and "What is the likelihood that Alice quantized her channel measurement to a `1', given Bob's channel measurement?"  Our approach for capturing this information is provided in the following paragraphs.

As mentioned earlier, Alice and Bob obtain different values for the same channel sample, due to the independent noise corrupting their measurements. By using equations~(\ref{eq:alice_sample})~and~(\ref{eq:bob_sample}) in one dimension, we can define the noise variable based on the difference as follows:

\begin{align}
e[n] - f[n] = w_e[n] - w_f[n] = w'[n], \label{eq:channelNoise}
\end{align}

\hfill \break
where $e[n] \in \{ \operatorname{Re}(a[n]) , \operatorname{Im}(a[n]) \}$, $f[n] \in \{ \operatorname{Re}(b[n]) , \operatorname{Im}(b[n]) \}$, $w_e[n] \in \{ \operatorname{Re}(w_a[n]) , \operatorname{Im}(w_a[n]) \}$, $w_f[n] \in \{ \operatorname{Re}(w_b[n]) , \operatorname{Im}(w_b[n]) \}$, and $w'[n]\sim \mathcal{N}(0,\sigma^2_w)$.  $a[n]$, $b[n]$, $w_a[n]$, and $w_b[n]$ are all defined in equations~(\ref{eq:alice_sample})~and~(\ref{eq:bob_sample}).  Therefore, the LLR for a given channel sample $h_{AB}[n]$ can be calculated as

\begin{align}
LLR_{chan}[n] &= \ln\frac{\mbox{Pr}(e[n]=\text{`0'}|f[n])}{\mbox{Pr}(e[n]=\text{`1'}|f[n])}. \label{eq:chanLLR1}
\end{align}

\hfill \break
Equation~(\ref{eq:chanLLR1}) mathematically represents the idea of determining the ratio of the conditional probability that Alice quantized her channel measurement to a `0' bit at sample $n$, given that Bob received value $f$ at sample $n$ to the conditional probability that Alice quantized her channel measurement to a `1' bit at sample $n$, given that Bob received value $f$ at sample $n$.  Inserting the criteria for Alice's hard-decisions (i.e. $e[n]=\text{`0'}$ if $e[n]>0$ and $e[n]=\text{`1'}$ if $e[n]<0$), equation~(\ref{eq:chanLLR1}) becomes:

\begin{align}
LLR_{chan}[n] &= \ln\frac{\mbox{Pr}(e[n]>0|f[n])}{\mbox{Pr}(e[n]<0|f[n])}. \label{eq:chanLLR2}
\end{align}

\hfill \break
By subtracting Bob's received value $f$ at sample $n$ from both sides of Alice's hard-decision inequalities, equation~(\ref{eq:chanLLR2}) becomes:

\begin{align}
LLR_{chan}[n] &= \ln\frac{\mbox{Pr}(\{e[n]-f[n]>0-f[n]\}|f[n])}{\mbox{Pr}(\{e[n]-f[n]<0-f[n]\}|f[n])}. \label{eq:chanLLR3}
\end{align}

\hfill \break
Since $f[n]$ is now a part of the probabilities in equation~(\ref{eq:chanLLR3}), the probabilities are no longer conditional upon $f[n]$, and equation~(\ref{eq:chanLLR3}) reduces to:

\begin{align}
LLR_{chan}[n] &= \ln\frac{\mbox{Pr}(e[n]-f[n]>0-f[n])}{\mbox{Pr}(e[n]-f[n]<0-f[n])}. \label{eq:chanLLR4}
\end{align}

\hfill \break
Removing the now unnecessary $0$, equation~(\ref{eq:chanLLR4}) reduces to:

\begin{align}
LLR_{chan}[n] &= \ln\frac{\mbox{Pr}(e[n]-f[n]>-f[n])}{\mbox{Pr}(e[n]-f[n]<-f[n])}. \label{eq:chanLLR5}
\end{align}

\hfill \break
Substituting equation~(\ref{eq:channelNoise}) into equation~(\ref{eq:chanLLR5}), equation~(\ref{eq:chanLLR5}) now becomes:

\begin{align}
LLR_{chan}[n] &= \ln\frac{\mbox{Pr}(w'[n]>-f[n])}{\mbox{Pr}(w'[n]<-f[n])}. \label{eq:chanLLR6}
\end{align}

\hfill \break

\begin{figure}
\centering
\includegraphics[scale=1]{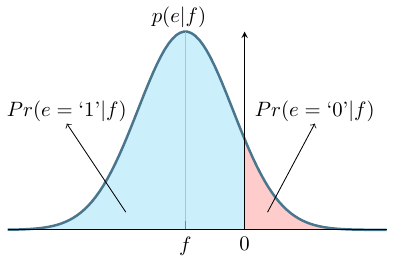}
\caption[Channel Measurement LLR Calculation Probabilities.]{\textbf{Channel Measurement LLR Calculation Probabilities.}}\label{fig:actual_llr}
\end{figure}

Since we know $w'[n]$ is a Gaussian random variable with known parameters, the corresponding probabilities of $e$ for known $f$ are calculated as illustrated in Figure~\ref{fig:actual_llr} by integrating the PDF on both sides of the $0$ quantization threshold separately to yield the LLR:

\begin{align}
LLR_{chan}[n] &= \ln \frac{\mathcal{Q} \left(\displaystyle \frac{-f[n]}{\sqrt{\sigma^2_w}} \right) }
{1-\mathcal{Q}\left(\displaystyle\frac{-f[n]}{\sqrt{\sigma^2_w}}\right)}.\label{eq:LLRq}
\end{align}

It should be noted for the example in Figure~\ref{fig:actual_llr} that $\mbox{Pr}(\cdot)$ indicates probability and $p(\cdot)$ indicates the value of the probability density function.  Also in Figure~\ref{fig:actual_llr}, $f$ is assumed to be negative but can take any real value.  The LLR calculation in equation~(\ref{eq:LLRq}) uses the $\mathcal{Q}$-function \cite{yates_goodman_prob}, which requires numerical integration or a stored lookup table.  To see how our proposed channel measurement LLRs function in our key generation process, observe the circle curve in Figure~\ref{fig:compareLLRs}.  At a minimum, the error that was generated at relatively high SNRs with the BPSK LLR approximation has successfully been eliminated by our proposed channel measurement LLRs.  A direct comparison of all our results will be presented in Chapter~\ref{sec:sims}.



%
%
%
%
%
%

\chapter{Low-Density Parity-Check Codes}\label{sec:LDPC}

Before presenting all of the comparison results of our simulations, we will explain some important details of LDPC codes that pertain to our work and, in particular, we will explain what a "code rate" means in the design for an error correction system.

\begin{figure}
\centering
\includegraphics[scale=1.25]{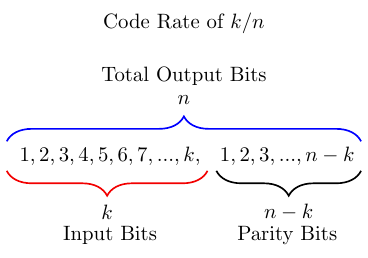}
\caption[Code Rate of $k/n$.]{\textbf{Code Rate of $k/n$.}}\label{fig:codeRate}
\end{figure}

In both communications and information theory, the "code rate" of an error-correcting code describes the proportion of the physically transmitted data that is considered to be both original information and non-redundant to the total amount of physically transmitted data.  The code rate is typically described as being $k/n$, meaning that for every $k$ input bits provided to the channel encoder, the channel encoder will output $n$ bits of data, of which $n-k$ are redundant parity bits \cite{FundErrorCodes}.  It is not uncommon for the parity bits to be generated and simply appended to the input bits in order to form the outputs.  A visual of this explanation is displayed in Figure~\ref{fig:codeRate}.

\begin{figure}
\centering
\includegraphics[scale=0.5]{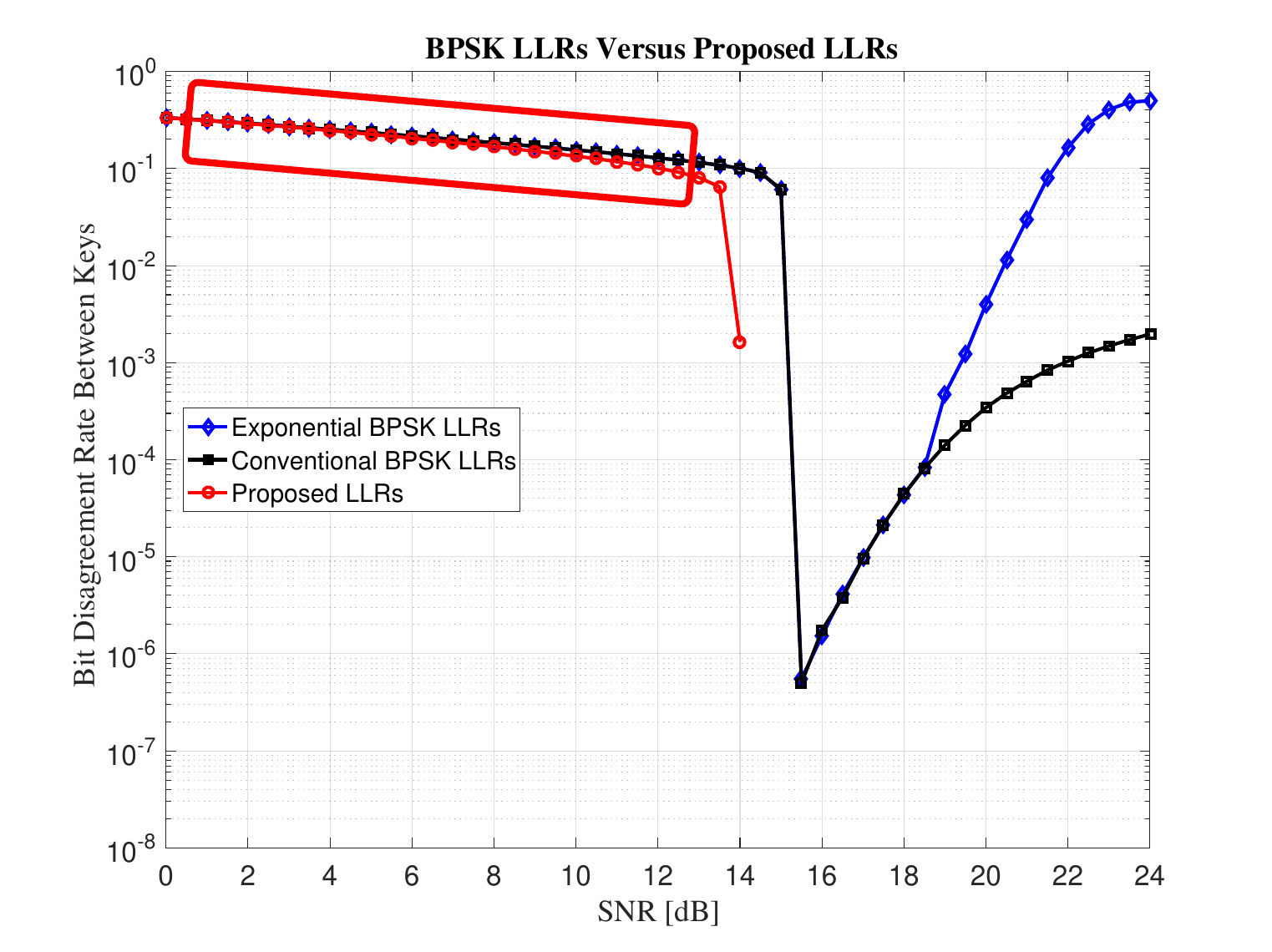}
\caption[The Flat Region of LDPC Codes.]{\textbf{The Flat Region of LDPC Codes.}}\label{fig:compareLLRs1}
\end{figure}

\begin{figure}
\centering
\includegraphics[scale=0.5]{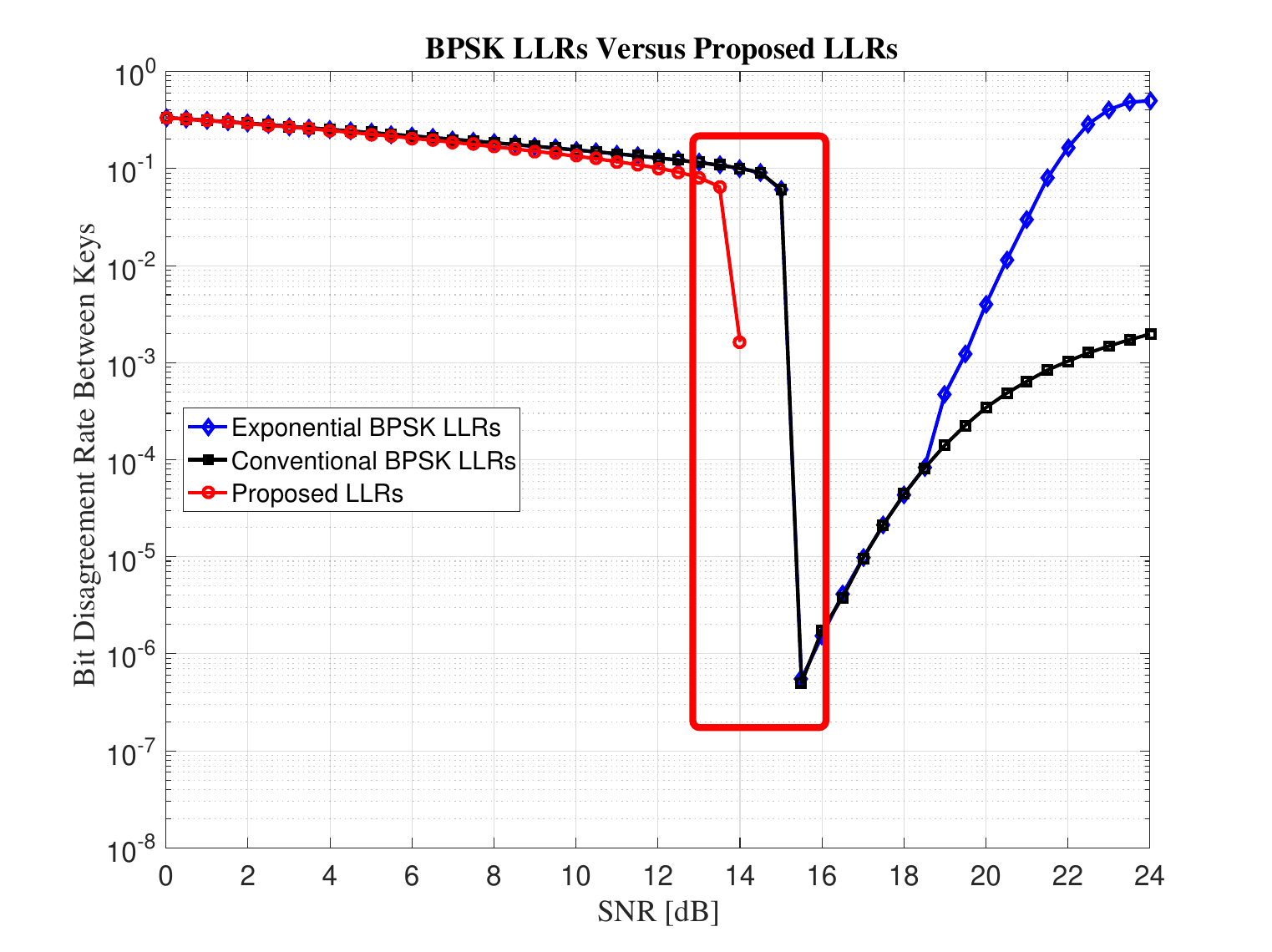}
\caption[The Waterfall Region of LDPC Codes.]{\textbf{The Waterfall Region of LDPC Codes.}}\label{fig:compareLLRs2}
\end{figure}

Other properties of LDPC codes that concern their BER plots are their \emph{flat region}, their \emph{waterfall region}, and their \emph{error-rate floor region} \cite{ChannelCodes, ErrorFloors}.  When observing the BER plots of LDPC codes, there are three distinctive regions that reveal themselves.  Going from low SNR to high SNR, the first region is relatively flat and comprises a high BER.  The error rate is generally too high in this first region and communications are not very effective.  The second region is termed the waterfall region.  The waterfall region is where the BER suddenly drops at a very significant rate, so steep that it has been named the waterfall region.  A simulation of the performance of a communication scheme employing a BPSK signaling scheme with LDPC codes for error correction was displayed in Figure~\ref{fig:compareLLRs}, and that plot's flat region and waterfall region have been highlighted in Figures~\ref{fig:compareLLRs1}~and~\ref{fig:compareLLRs2}.  As can be seen, the relatively flat region with high BER and the waterfall regions of the three curves in Figures~\ref{fig:compareLLRs1}~and~\ref{fig:compareLLRs2} exemplify why they are termed as such.  The third region of LDPC code BER plots, which cannot be seen in Figure~\ref{fig:compareLLRs}, is the error-rate floor region.  The slope of the error-rate floor region is not necessarily as flat as the slope of the first region; however, the BER no longer drops as steeply as in the waterfall region.  The reason the error-rate floor region is not displayed in Figure~\ref{fig:compareLLRs} is that it can take a very long time for enough simulations to be performed in order to even find where the error-rate floor region begins.  Sometimes, these simulations can require months of running before a BER can be determined, because the error rate is so low \cite{ChannelCodes, ErrorFloors}.  The waterfall region will comprise most of the comparisons made for our simulations.

One of the prevailing reasons for these waterfall regions in LDPC codes has to do with approaching the theoretical Shannon limit of a communication system for a given bandwidth and SNR.  Due to the fact that our key generation mechanism is not a typical communication system, we do not prove or show the theoretical Shannon limit of our system.  However, in a BPSK communication system that uses LDPC codes for correcting errors, the theoretical Shannon limit is often drawn on the BER plot of the BPSK communication system to display how close the LDPC encoding brings the communication system to the Shannon limit.  A few examples can be seen on on pages 436--459 of Reference~\cite{ChannelCodes}.  The figures in those pages display the BERs of a BPSK communication system with a specified bandwidth, along with the Shannon limit for that BPSK system.  Also displayed in those figures are one or more curves revealing how a given LDPC code improves the BER performance of the BPSK communication system.

In plots that reveal the BER performance improvement of LDPC codes, the best designed LDPC codes are shown to begin their waterfall region very shortly after the Shannon limit.  The ability of a well-designed LDPC code to allow a communication system to approach the system's theoretical limit gives its BER plot a waterfall region because the Shannon limit is a vertical line.  The ability that LDPC codes have to help bring a communication system close to the Shannon limit is a prevailing reason why LDPC codes are studied.  For additional information concerning LDPC codes, please see the following references:  \cite{Forney1, waterfall2016, messagePassing, RobertGallagerLDPC, Efficient, Erasure, Lowering, Expander, TurboLike, Irregular, Good, MinDistTurbo, RepeatAccum, Recursive}.


\chapter{Simulations and Numerical Results}\label{sec:sims}
In this chapter, we present numerical results obtained from Monte-Carlo simulations to evaluate the probability of bit disagreement between the two secret keys $\kv_A$ and $\kv_B$ generated by Alice and Bob, respectively, for the proposed key generation and reconciliation scheme.  The simulations were performed for a range of SNR values and a range of LDPC code rates.  We note that regular LDPC codes from the DVB-S.2 standard \cite{DVB_LDPC} have been used for generating the side information.  We also compare using the one-bit quantization channel estimation (non-censoring) method with the channel estimation censoring scheme approach \cite{SecretKeyReactance, Cens_mil}, with the censoring threshold set to $\gamma=\sigma^2_{h_{BA}}/10$.

For both the one-bit quantization method and the censoring scheme for secret key generation, we used an LDPC code with a block length of $64$,$800$ to generate the side information.  Also, for each SNR value, the simulations were repeated for $3$,$000$ blocks.  Both BPSK approximation LLR calculations and our proposed LLR calculations were performed at Bob, when Bob was quantizing the channel in order to compare their effects on the secret key reconciliation process.  For the approximate LLR calculation, the amplitude of the BPSK signal in equation~(\ref{eq:approxLLR6}) was set to $\sqrt{E}=1$.  In all of the simulations, while transmitting side information from Alice to Bob via BPSK transmission, the rate of the LDPC code used for protecting the side information from transmission errors was fixed to $1/2$.

\begin{figure}
\centering
\includegraphics[scale=0.5]{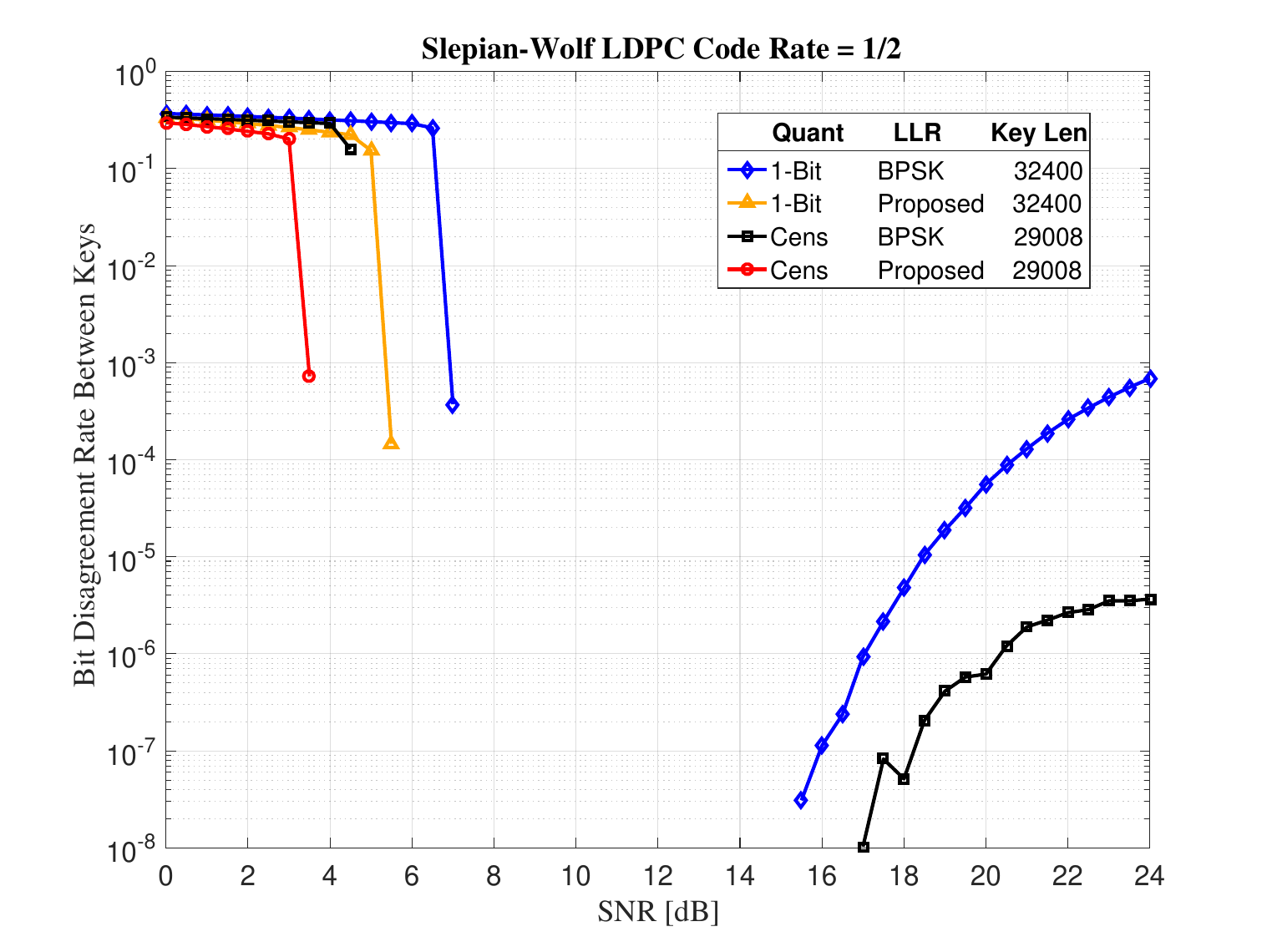} 
\caption[Simulation Results for Slepian-Wolf LDPC Code Rate of $1/2$.]{\textbf{Simulation Results for Slepian-Wolf LDPC Code Rate of $1/2$.}}\label{fig:rate12}
\end{figure}

\begin{figure}
\centering
\includegraphics[scale=0.5]{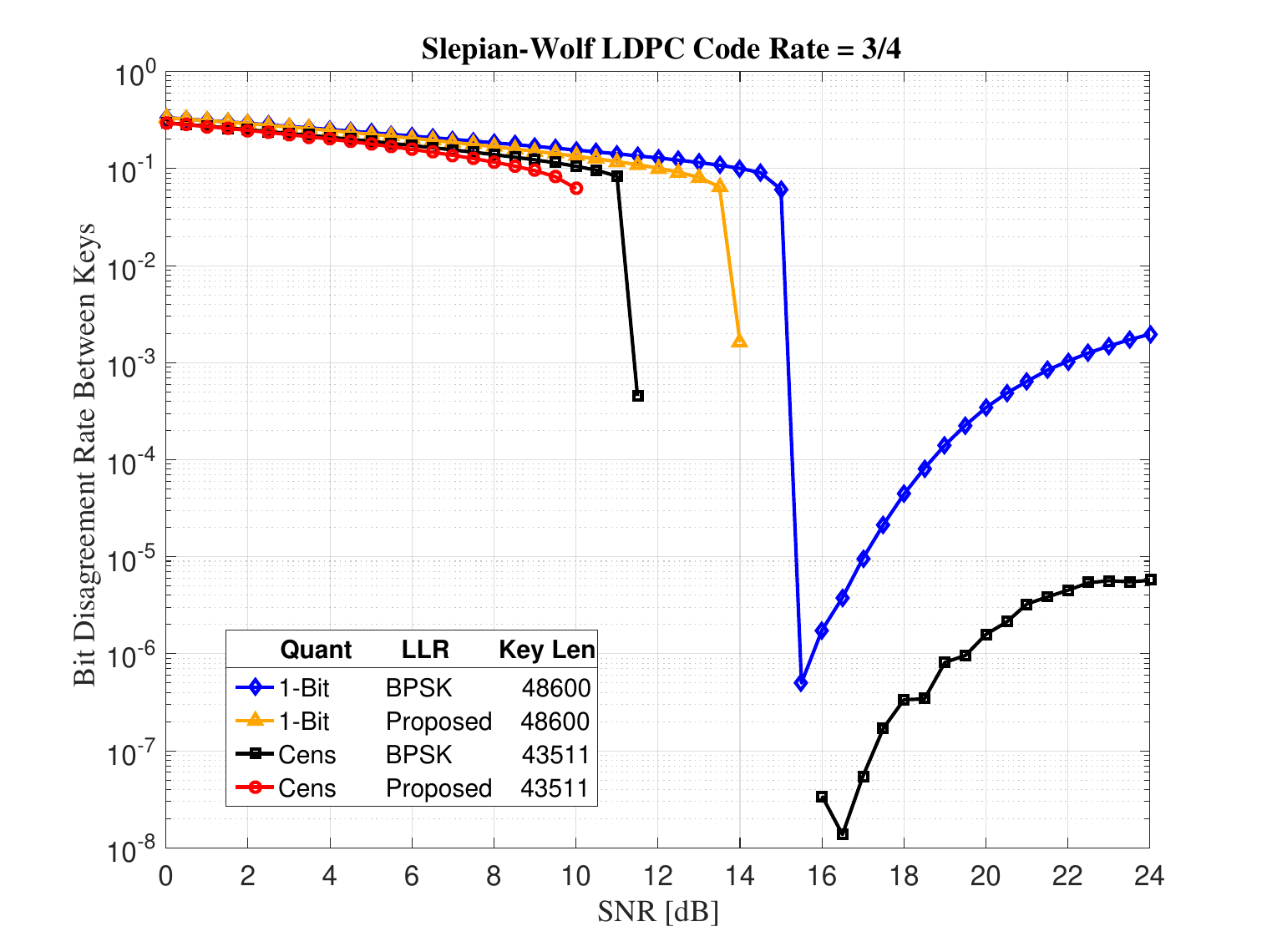} 
\caption[Simulation Results for Slepian-Wolf LDPC Code Rate of $3/4$.]{\textbf{Simulation Results for Slepian-Wolf LDPC Code Rate of $3/4$.}}\label{fig:rate34}
\end{figure}

\begin{figure}
\centering
\includegraphics[scale=0.5]{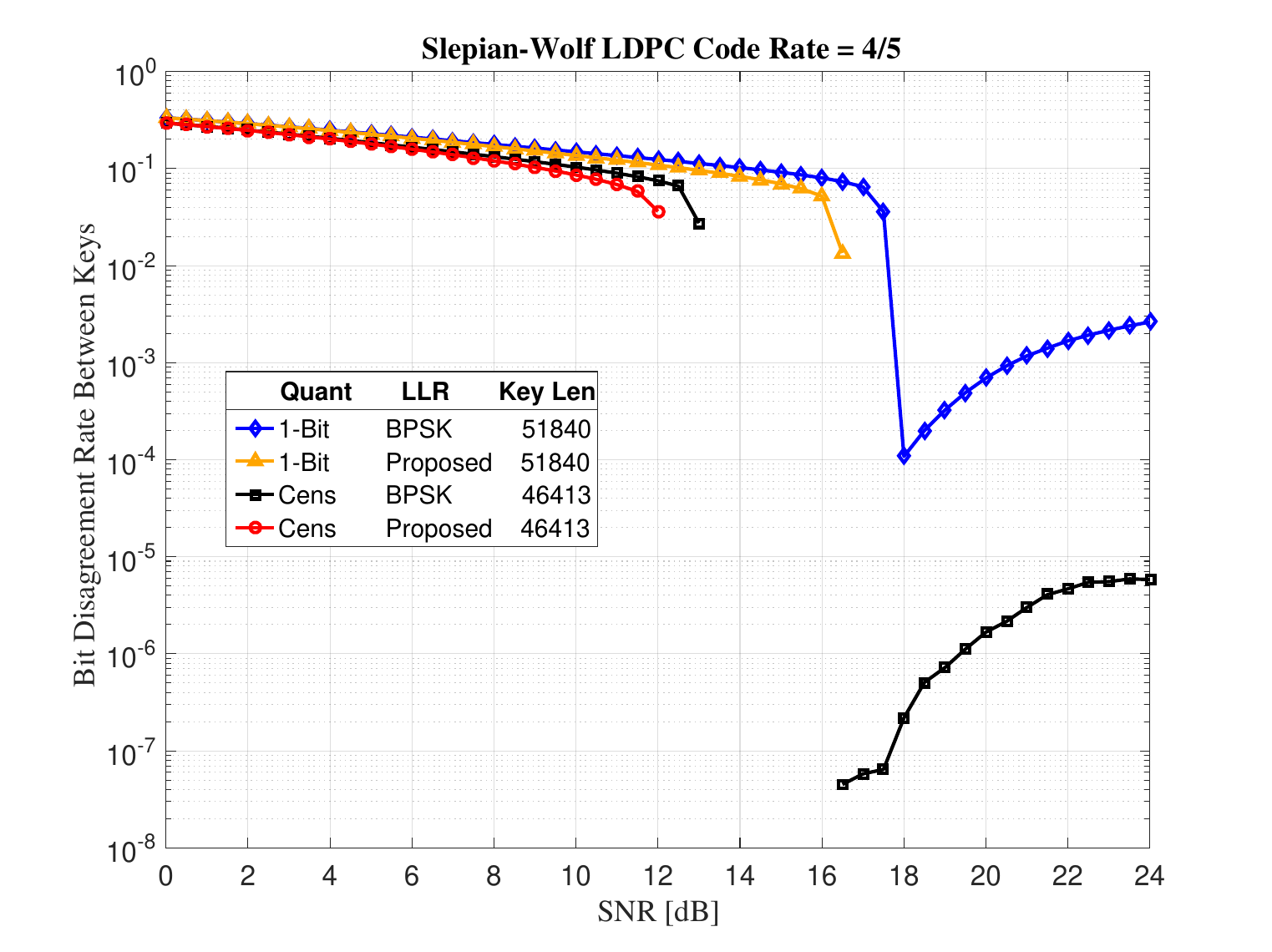} 
\caption[Simulation Results for Slepian-Wolf LDPC Code Rate of $4/5$.]{\textbf{Simulation Results for Slepian-Wolf LDPC Code Rate of $4/5$.}}\label{fig:rate45}
\end{figure}

\begin{figure}
\centering
\includegraphics[scale=0.5]{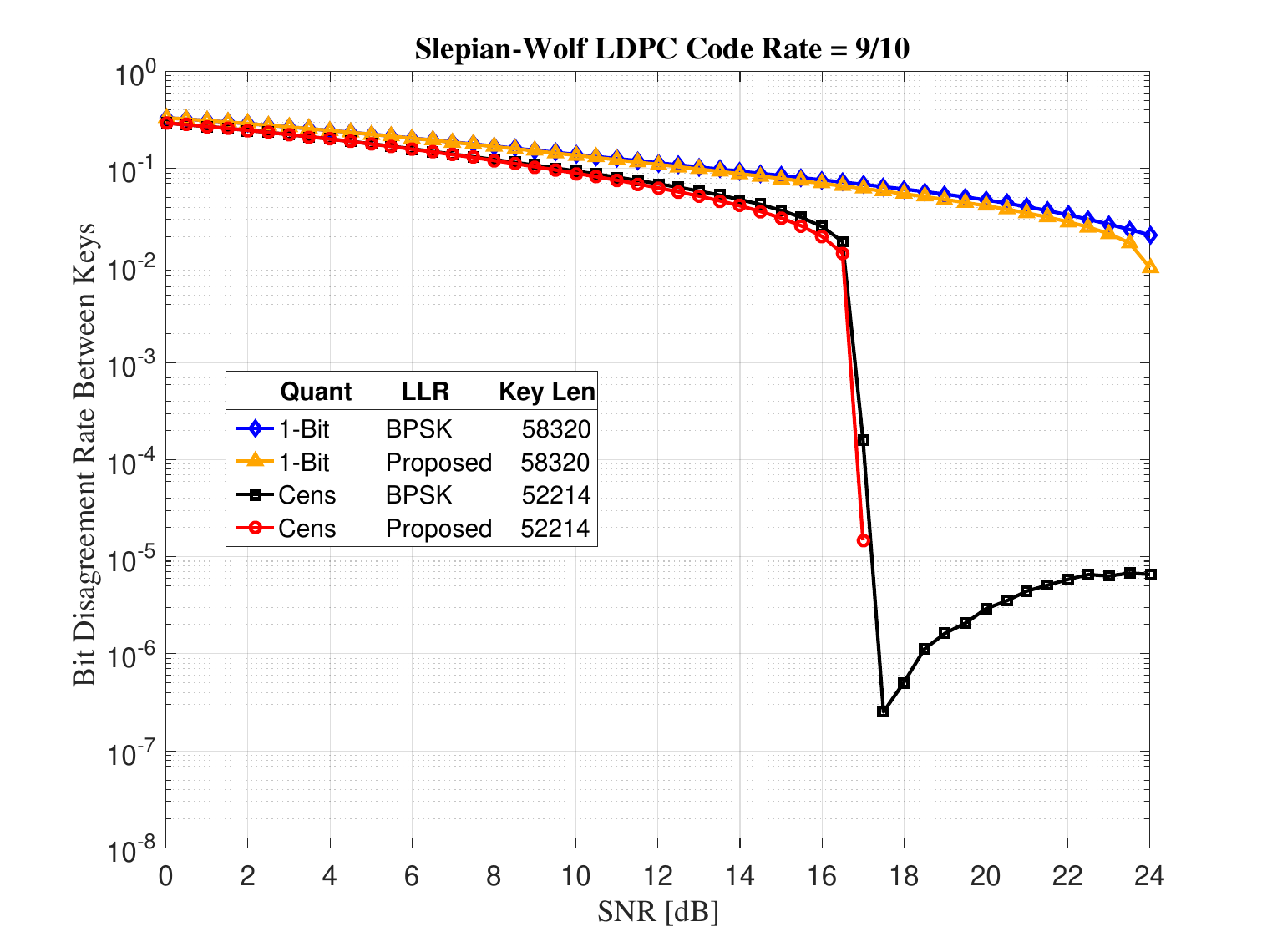} 
\caption[Simulation Results for Slepian-Wolf LDPC Code Rate of $9/10$.]{\textbf{Simulation Results for Slepian-Wolf LDPC Code Rate of $9/10$.}}\label{fig:rate910}
\end{figure}

Before discussing the simulation results shown in Figures~\ref{fig:rate12},~\ref{fig:rate34},~\ref{fig:rate45},~and~\ref{fig:rate910}, it is important to note a few facts from those figures and their legends, in order to make it easier to digest the information presented in the simulation result figures.  The curve with diamonds represents the one-bit quantization scheme being used with BPSK LLR calculations; we will refer to this as the "diamond curve."  The curve with triangles represents the one-bit quantization scheme being used with our proposed LLR calculations; we will refer to this as the "triangle curve."  The curve with squares represents the censoring scheme being used with BPSK LLR calculations; we will refer to this as the "square curve."  The curve with circles represents the censoring scheme being used with our proposed LLR calculations; we will refer to this as the "circle curve."

As we will see in the paragraphs that follow, the diamond curve can be compared with the triangle curve, and the square curve can be compared with the circle curve, in order to compare the LLR calculation methods.  Also, the diamond curve can be compared to the square curve, and the triangle curve can be compared to the circle curve, in order to compare the one-bit quantization scheme to the censoring scheme.  The triangle curve cannot be directly compared to the square curve and the diamond curve cannot be directly compared to the circle curve, because those comparisons would involve more than one changing variable.  Each figure can, however, be compared as a whole to the other figures to see the effects of increasing or decreasing the Slepian-Wolf LDPC code rate.

The simulation results, shown in Figures~\ref{fig:rate12},~\ref{fig:rate34},~\ref{fig:rate45},~and~\ref{fig:rate910}, provide insight on the effect of LDPC code rates, the effect of censoring, and, most importantly to our work, the effect of the BPSK approximate LLR calculations compared to our proposed LLR calculations.

There are several comparisons that can be made from observing Figures~\ref{fig:rate12}--\ref{fig:rate910}, and a summary of those comparisons is listed in Table~\ref{table:1}.  The first comparison that will be considered is actually a comparison between figures and specifically where their waterfall regions begin.

\setlength{\arrayrulewidth}{1mm}

\begin{table}
\centering
\caption[Trade-Offs:  SNR for Security.]{\textbf{Trade-Offs:  SNR for Security.}}
{
\rowcolors{2}{white}{cyan!20}
\begin{tabular}{ |M{0.4in}|R{0.75in}|R{0.82in}|R{0.765in}|R{0.905in}|R{0.7in}|R{0.65in}|  }
\hline
\rowcolor{green!20}
\textbf{\newline \newline \newline \newline \newline Code Rate} & \centering\textbf{\newline \newline Waterfall Region Starting SNR [dB]} & \centering\textbf{\newline Gap Between Censoring and Non-Censoring [dB]} & \centering\textbf{Gap Between BPSK and Proposed LLRs [dB]} & \centering\textbf{\newline \newline \newline \newline Key Length (Censored)} & \centering\textbf{\newline \newline \newline \newline Revealed Parity Bits} & \textbf{\newline \newline \newline \newline \newline \newline Security} \\
\hline
$1/2$  & $3.0$  & $2.0$ & $1.5$ & $32$,$400$ ($29$,$008$) & $32$,$400$ & Least Secure \\
$3/4$  & $10.0$ & $4.0$ & $1.5$ & $48$,$600$ ($43$,$511$) & $16$,$200$ & Less Secure \\
$4/5$  & $12.0$ & $4.5$ & $1.0$ & $51$,$840$ ($46$,$413$) & $12$,$960$ & More Secure \\
$9/10$ & $16.5$ & $7.5$ & $0.5$ & $58$,$320$ ($52$,$214$) & $6$,$480$  & Most Secure \\
\hline
\end{tabular}
\label{table:1}
}
\end{table}

In Figure~\ref{fig:rate12} with a Slepian-Wolf code rate of $1/2$, the waterfall regions of the four curves begin at an SNR as low as 3~dB (circle curve) to as high as 6.5~dB (diamond curve).  In Figure~\ref{fig:rate34} with a Slepian-Wolf code rate of $3/4$, the waterfall regions of the four curves begin at an SNR as low as 10~dB (circle curve) to as high as 15~dB (diamond curve).  In Figure~\ref{fig:rate45} with a Slepian-Wolf code rate of $4/5$, the waterfall regions of the four curves begin at an SNR as low as 12~dB (circle curve) to as high as 17.5~dB (diamond curve).  In Figure~\ref{fig:rate910} with a Slepian-Wolf code rate of $9/10$, the waterfall regions of the four curves begin at an SNR as low as 16.5~dB (circle curve) to as high as greater than 24~dB (diamond curve).

The changes of SNR values where the waterfall regions begin intuitively make sense when we consider the effect that the code rate has on the key generation and reconciliation process.  Since our LDPC codes function by using size $64$,$800$ bit-length blocks, the total number of secret key bits plus the total number of side information bits must equal $64$,$800$.  Thus, for a Slepian-Wolf code rate of $1/2$, there are $32$,$400$ secret key bits and $32$,$400$ revealed parity bits.  For a Slepian-Wolf code rate of $3/4$, there are $48$,$600$ secret key bits and $16$,$200$ revealed parity bits.  For a Slepian-Wolf code rate of $4/5$, there are $51$,$840$ secret key bits and $12$,$960$ revealed parity bits.  For a Slepian-Wolf code rate of $9/10$, there are $58$,$320$ secret key bits and $6$,$480$ revealed parity bits.

With the Slepian-Wolf code rate being representative of the number of secret key bits to the total number of bits, and by observing Figures~\ref{fig:rate12}--\ref{fig:rate910}, it is easy to understand the tradeoff that occurs when changing the rate of the Slepian-Wolf code.  The lower the Slepian-Wolf code rate, the fewer the number of secret key bits that are generated, and the lower the SNR can be for successful key agreement for Alice and Bob; however, with fewer secret key bits generated, more side information bits are generated, and therefore more information about the secret key is revealed, lowering the security of the secret key.  Essentially, with a low Slepian-Wolf code rate comes lessened security but a lower minimum SNR requirement; also, with a high Slepian-Wolf code rate comes greater security, but a higher minimum SNR requirement.  Clearly, there is a tradeoff between security and minimum required SNR.

The second comparison that we will concentrate on is the effect that censoring has on the key generation process.  By looking at Figures~\ref{fig:rate12},~\ref{fig:rate34},~\ref{fig:rate45},~and~\ref{fig:rate910} individually and by comparing the appropriate curves to one another (i.e. the diamond curve to the square curve and the triangle curve to the circle curve), you can see that the censoring scheme with its threshold set to $\gamma=\sigma^2_{h_{BA}}/10$ always provides an improvement in SNR for where the waterfall region begins.

In Figure~\ref{fig:rate12} with a Slepian-Wolf code rate of $1/2$, comparing the triangle curve to the circle curve, censoring lowered the SNR requirement of where the waterfall region begins by about 2~dB.  Comparing the diamond curve to the square curve, censoring once again lowered the SNR requirement of where the waterfall region begins by about 2~dB.  In Figure~\ref{fig:rate34} with a Slepian-Wolf code rate of $3/4$, comparing the triangle curve to the circle curve, censoring lowered the SNR requirement of where the waterfall region begins by about 3.5~dB.  Comparing the diamond curve to the square curve, censoring lowered the SNR requirement of where the waterfall region begins by about 4~dB.  In Figure~\ref{fig:rate45} with a Slepian-Wolf code rate of $4/5$, comparing the triangle curve to the circle curve, censoring lowered the SNR requirement of where the waterfall region begins by about 4.5~dB.  Comparing the diamond curve to the square curve, censoring once again lowered the SNR requirement of where the waterfall region begins by about 4.5~dB.  In Figure~\ref{fig:rate910} with a Slepian-Wolf code rate of $9/10$, comparing the both the triangle curve to the circle curve and the diamond curve to the square curve, censoring lowered the SNR requirement of where the waterfall region begins by at least 7.5~dB.

You can see by looking at the legends in Figures~\ref{fig:rate12}--\ref{fig:rate910} that the number of secret key bits generated using the censoring quantization scheme is about $10\%$ less than the number of secret key bits generated using the one-bit quantization scheme (look for the "key length" column in the figure legends).  The tradeoff presented in the comparison is once again security for the minimum SNR requirement.  For every bit that Alice censors, Alice must transmit the bit locations that she censored to Bob, revealing to any eavesdroppers which bits from the secret key can be thrown away.  Granted, this information is most likely less revealing than transmitting the same number of parity bits, since the side information allows eavesdroppers to potentially correct key bits rather than just to throw key bits away.  Also, the censoring threshold is easier to control and adjust than the Slepian-Wolf LDPC code rate, so perhaps combining a balance of the Slepian-Wolf code rate and the censoring threshold may prove to be the best formulation, in the future.

The final comparison that we will make, and the most important to our work, is the improvement in bit disagreement rate between the secret keys of Alice and Bob using our proposed LLR calculations instead of the BPSK approximation LLR calculations.  One of the first irregularities that you may notice in Figures~\ref{fig:rate12}--\ref{fig:rate910} occurs at SNR values greater than about 15~dB.  The diamond and square curves, representative of the simulations using the BPSK approximation LLR calculations, do not exhibit typical LDPC coded BER results after the SNR increases beyond about 15~dB.  Instead, their bit disagreement rates begin to steadily increase as the SNR continues to increase.  This increase in bit disagreement rate is due to the fact that the BPSK approximation is no longer a good approximation for our situation.  The BPSK approximation is no longer true at SNR values greater than about 15~dB and thus produces errors.

When you observe the triangle and circle curves, representative of the simulations using our proposed LLR calculations, you notice that the curves appear to simply stop at the waterfall region instead of continuing on.  This is due to the fact that, at those SNR values and greater, there were no errors that occurred within the number of simulations that we ran; therefore, the logarithmic bit disagreement rate scale did not plot any results, because they would be $-\infty$, which is not a number.

Therefore, at a minimum, our proposed LLR calculations are required to be used in order to allow the system to continue to function at SNR values beyond the waterfall regions.  There is, however, another comparison between the BPSK approximation LLR calculations and our proposed LLR calculations.

In Figure~\ref{fig:rate12}, comparing the diamond curve to the triangle curve, our proposed LLR calculations provide an approximate 1.5~dB gain in SNR over the BPSK approximation LLR calculations.  Comparing the square curve to the circle curve, our proposed LLR calculations also provide an approximate 1.5~dB gain in SNR over the BPSK approximation LLR calculations.  In Figure~\ref{fig:rate34}, comparing the diamond curve to the triangle curve, our proposed LLR calculations provide an approximate 1.5~dB gain in SNR over the BPSK approximation LLR calculations.  Comparing the square curve to the circle curve, our proposed LLR calculations provide an approximate 1~dB gain in SNR over the BPSK approximation LLR calculations.  In Figure~\ref{fig:rate45}, comparing the diamond curve to the triangle curve, our proposed LLR calculations provide an approximate 1~dB gain in SNR over the BPSK approximation LLR calculations.  Comparing the square curve to the circle curve, our proposed LLR calculations also provide an approximate 1~dB gain in SNR over the BPSK approximation LLR calculations.  In Figure~\ref{fig:rate910}, comparing the diamond curve to the triangle curve, our proposed LLR calculations provide less than an approximate 0.5~dB gain in SNR over the BPSK approximation LLR calculations.  Comparing the square curve to the circle curve, our proposed LLR calculations also provide less than an approximate 0.5~dB gain in SNR over the BPSK approximation LLR calculations.

By comparing the gain in SNR of our proposed LLR calculations to the BPSK approximation LLR calculations, you can see that our proposed LLR calculations provide more gain with lower Slepian-Wolf LDPC code rates (i.e. $1/2$ and $3/4$) than they do with higher Slepian-Wolf LDPC code rates (i.e. $4/5$ and $9/10$).  In the worst-case scenario, with a Slepian-Wolf LDPC code rate of $9/10$, the error that arises from the BPSK approximation is eliminated but the gain in SNR is only very slight.  In the best-case scenario, with a Slepian-Wolf LDPC code rate of $1/2$, the error that arises from the BPSK approximation is also eliminated, and the gain in SNR is 1.5~dB.

In all of the simulated Slepian-Wolf code rate scenarios, and even in the worst-case scenario, the error that arose from the BPSK approximation was eliminated by using our proposed LLR calculations instead, which became a necessary and useful improvement to the physical layer secret key agreement scheme.

\chapter{Conclusion}\label{sec:con}
In this thesis, we discussed a brief overview of cryptography and we provided a detailed description of physical layer secret key agreement using channel state information.  Advantage distillation or key generation was explained, information reconciliation or secret key reconciliation was described, and privacy amplification or cryptographic hashing was mentioned.  We explained how logarithmic-likelihood ratios (LLRs) can be used to form soft-decisions, proved how Binary Phase-Shift Keying (BPSK) LLRs are optimally calculated from received BPSK transmissions, and even used BPSK LLRs for approximations when calculating key reconciliation LLRs.  We also showed how we derived our key reconciliation LLRs.  We gave a brief explanation of Low-Density Parity-Check (LDPC) codes to explain how they are important to key reconciliation and to explain how to discern and interpret their bit error rate (BER) plots.  Finally, we presented our simulations and numerical results depicting how our proposed LLRs outperform the BPSK approximation LLRs in the purpose of key reconciliation.

\section{Results}\label{sec:conRes}
To be specific, the thesis presents a new method for secret key generation that is based on one-bit quantization of channel measurements along with LLRs calculated using the difference between channel estimates at legitimate reciprocal nodes.  The proposed approach eliminates the need for complex LLR calculations required by key generation based on vector quantization of channel measurements \cite{LDPC_code_const, LDPC_Quant} while also implementing advantage distillation, along with information reconciliation using Slepian-Wolf LDPC codes.

Numerical results are also presented, in which the one-bit quantization scheme is compared to a censoring scheme for key generation that also uses channel state information, with exact and approximate calculations for LLRs.  Both schemes were simulated for different SNR values and for different Slepian-Wolf LDPC code rates.  Our results confirm that the probability of bit disagreement between the keys generated by Alice and Bob decreases with increasing SNR, reaching very low values after the waterfall region.  The SNR values of the waterfall regions depend on the rate of the Slepian-Wolf LDPC code employed.

\newpage

\section{Future Directions}\label{sec:conDir}
Generating secret keys from the inherent randomness of wireless channels is a promising technique to improve the sharing of cryptographic keys between legitimate users and to speed up the key agreement process over that of public key cryptosystems.  However, there are still open areas that must be resolved in order to make secret key generation more robust and practical.  Several future research agendas are summarized below:

\begin{enumerate}
\item Secret key generation in static environments.  Although researchers have attempted to introduce randomness into static environments with long channel coherence times using parasitic antenna arrays such as the RECAP \cite{recap2012} or ESPAR \cite{espar2003} arrays, and while these methods do decrease the channel coherence time, it is not necessarily practical to use reconfigurable antenna arrays in many products.  Approaches that utilize multiple-input multiple-output (MIMO) antenna arrays may be more desirable, since they are easier to implement in practice \cite{MIMOfuture1}.  The ability to perform secret key agreement in a static environment is essential for the practical application and installation of secret key generation systems.
\item Attacks against secret key generation systems.  Research into attacking secret key generation systems currently receives limited input.  It has been indicated that key generation is vulnerable to passive eavesdropping \cite{PassiveAttacks, InviteTheThief}, to active attacks \cite{ActiveAttackers, InjectionAttacks}, and even to a combination of the two \cite{RobustKeyExtraction}.  Research that looks into how we can undermine such attacks, or fortify against these attacks, is necessary if we are to construct practical, durable, and secure key generation systems.
\end{enumerate}


\bibliographystyle{IEEEtran}
\renewcommand\bibname{\uppercase{Bibliography}}
\cleardoublepage
\phantomsection
\addcontentsline{toc}{chapter}{\uppercase{Bibliography}}
\bibliography{./bib/refs}
%
%
%

\vitapage

\end{document}